\documentclass[12pt]{article}

\usepackage[a4paper,left=30mm,right=20mm,top=20mm,bottom=20mm]{geometry}
\usepackage{graphics}
\usepackage{amsmath}
\usepackage{epsfig}
\usepackage{cite}
\usepackage{xcolor}
\usepackage[unicode]{hyperref}
\newcommand{\bea}{\begin{eqnarray}}
\newcommand{\eea}{\end{eqnarray}}
\newcommand{\be}{\begin{equation}}
\newcommand{\ee}{\end{equation}}

\title{Updating TMD parton densities in a proton within the Kimber-Martin-Ryskin approach}

\author{A.V.~Kotikov$^{1}$, A.V.~Lipatov$^{2}$}

\begin{document}

\maketitle

\begin{center}
{\it $^{1}$Joint Institute for Nuclear Research, 141980 Dubna, Moscow region, Russia}\\
{\it $^{2}$Skobeltsyn Institute of Nuclear Physics, Lomonosov Moscow State University, 119991 Moscow, Russia\\}

\end{center}

\vspace{0.5cm}

\begin{center}

{\bf Abstract }

\end{center}

\indent
We present analytical expressions for the Transverse Momentum Dependent (TMD, or unintegrated) gluon
and quark densities in a proton derived at leading order of QCD running coupling
and valid at both small and large $x$.
The calculations are performed
using the
%fixed-flavor-number scheme using the
Kimber-Martin-Ryskin/Watt-Martin-Ryskin %(KMR)
prescription (in both differential and integral formulations)
with different treatment of kinematical constraint, which reflects the angular and
strong ordering conditions for parton emissions.
As an input, analytical solution of QCD evolution equations
for conventional (collinear)
parton distributions is applied,
%at small and large $x$ values,
where the valence and non-singlet quark parts obey the Gross-Llewellyn-Smith and
Gottfried sum rules %respectively
and momentum conservation for the singlet quark and gluon densities is taken into account.
Several phenomenological parameters are extracted from combined fit to precision
data on the proton structure function $F_2(x,Q^2)$ collected
by the BCDMS, H1 and ZEUS Collaborations,
comprising a total of $933$ points from $5$ data sets.
Comparison with the numerical results
obtained by other groups is presented
and
%some
phenomenological application %of the derived TMD parton densities
to the inclusive $b$-jet production at the LHC is given.

\vspace{1.0cm}

\noindent{\it Keywords:} QCD evolution, parton density functions, small-$x$ physics, Kimber-Martin-Ryskin approach, High Energy Factorization

%\newpage
\vspace{1.0cm}

\newpage

\section{Introduction} \indent

It is well known that theoretical description of hadronic collisions at high energies %processes at hadron colliders
which proceed with large momentum transfer and involve several hard scales can be
achieved in the framework of High Energy Factorization\cite{HighEnergyFactorization} (or $k_T$-factorization\cite{kt-factorization})
approach of Quantum Chromodynamics (QCD). An essential point of this formalism %the latter
is using
the Transverse Momentum Dependent (TMD) parton (gluon
and quark) distribution functions\footnote{In the $k_T$-factorization literature,
the TMD parton densities are also often referred
as unintegrated parton distributions (uPDFs).
Hereafter we will use first notation, TMDs.} in a proton and/or nuclei,
$f_a(x, {\mathbf k}_T^2, \mu^2)$.
These quantities
depend on the fraction $x$ of the colliding hadron longitudinal momentum carried by an
interacting parton $a$, its two-dimensional transverse momentum ${\mathbf k}_T^2$ and hard scale $\mu^2$
of the corresponding partonic scattering subprocess.
The TMD parton %(gluon and quark)
densities encode nonperturbative information on hadron structure,
including transverse momentum and polarization degrees of freedom
and are under active investigation at present\footnote{They are
widely used also in the Collins-Soper-Sterman approach (or TMD factorization)\cite{CSS-1,CSS-2,CSS-3,CSS-4},
which is designed for semi-inclusive processes with a finite and non-zero ratio
between the hard scale $\mu^2$ and total energy $s$
(see also reviews\cite{TMD-review1,TMD-review2} for more information).}.
In particular, their detailed knowledge is necessary for accurate theoretical
predictions for cross sections of number of semi-inclusive
and multiscale QCD processes studied currently at modern colliders (LHC, RHIC) and
planned to be studied at future machines (FCC-he, EiC, EicC, NICA, CEPC etc).

In contrast with a great amount of our knowledge about the conventional (collinear)
parton distribution functions (PDFs), $f_a(x,\mu^2)$, accumulated
%in theoretical and experimental studies
over past years, the
TMD parton densities
in a proton and especially nuclei
are rather
poorly known quantities still.
In the asymptotical limit of high energies (or, equivalently, low $x$)
the TMD gluon distributions %in a proton
must satisfy famous Balitsky-Fadin-Kuraev-Lipatov (BFKL)\cite{BFKL}
%Catani-CiafaloniFiorani-Marchesini (CCFM)\cite{?}
evolution equation, where
large logarithmic contributions to the
production cross section proportional
to $\alpha_s^n \ln^n s/\Lambda_{\rm QCD}^2 \sim \alpha_s^n \ln^n 1/x$
are taken into account.
Such terms
are expected to be even more important in comparison with conventional
Dokshitzer-Gribov-Lipatov-Altarelli-Parisi (DGLAP)\cite{DGLAP} contributions, which are proportional to $\alpha_s^n \ln^n \mu^2/\Lambda_{\rm QCD}^2$.
Also, the Catani-Ciafaloni-Fiorani-Marchesini (CCFM) equation\cite{CCFM}, which additionally
resums terms proportional to $\alpha_s^n \ln^n 1/(1 - x)$
and BK\cite{BK} or JIMWLK equations\cite{JIMWLK-1,JIMWLK-2,JIMWLK-3,JIMWLK-4,JIMWLK-5,JIMWLK-6}, where non-linear effects
are taken into account, can be used to
describe the non-collinear (dependent on the transverse momentum) QCD evolution.

Other popular approaches to derive
the TMD parton distributions are the so-called
Kimber-Martin-Ryskin (KMR)\cite{KMR-LO} or Watt-Martin-Ryskin (WMR)\cite{WMR-LO} prescriptions, which
are formalisms to construct the TMDs
from well-known collinear %(collinear) PDFs.
PDFs.
Note that the WMR approach is an extension and further development
of early KMR ideas and
%The KMR approach is explored currently at leading order (LO)\cite{KMR-LO} and next-to-leading order (NLO)\cite{KMR-NLO}.
explored currently at the next-to-leading order (NLO)\cite{KMR-NLO}.
The key assumption of the KMR/WMR formalism is
that the dependence of parton distributions on the transverse momentum %dependence %of the parton densities
enters only at the last evolution step, so
that the standard DGLAP %evolution
scenario could be applied up to
this step. %\footnote{Another DGLAP-based scheme to evaluate the TMD parton densities in a proton
%is the Parton Branching (PB) approach\cite{PB1,PB2}.}.
%The KMR approach is currently explored at the next-to-leading order (NLO)\cite{?}
%and widely used in the phenomenological applications (see, for example,\cite{?}).
Such prescription, where
ordinary PDFs (as obtained numerically, for example, by the
NNPDF\cite{NNPDF4}, MHST\cite{MSHT20}, CTEQ-TEA\cite{CT14, CT18} or IMP\cite{IMP} groups)
are employed as an input for the KMR/WMR procedure,
is widely used in the phenomenological applications (see, for example,\cite{Szczurek-HFjj,Szczurek-phi,Motyka-photon,Iran-WZ,KMR-VFNS-HF1,KMR-VFNS-HF2} and
references therein).

In the present paper, the KMR/WMR formalism is employed for analytical
calculations
%(see, for example,\cite{?} and references therein).}
of the TMD parton densities in a proton
valid at both small and large $x$.
As a first step,
our calculations are limited to leading order (LO) of the perturbation theory.
The LO approximation is reasonable for processes, where the NLO corrections
are still not known. Moreover, most of phenomenological applications of
$k_T$-factorization approach %involving TMDs
are currently performed at LO\footnote{First attempts
to include higher order corrections were done (see, for example,\cite{NLO-kt-cjets,NLO-kt-charmonia-our} and references therein).}.
Our derivation is mainly based on the
expressions\cite{PDFs-our} for conventional PDFs
obtained (in the fixed-flavor-number scheme, FFNS) from the analytical solution of the DGLAP evolution equations.
%which are used as an input in the KMR/WMR procedure.
%procedure.
The expressions\cite{PDFs-our} rely %have origin
on exact
asymptotics at low and large values of $x$ and
contain subasymptotic terms fixed by the momentum conservation and, in the non-singlet and
valence parts, by the Gross-Llewellyn-Smith and Gottfried sum rules, respectively.
In contrast with the preliminary study\cite{PDFs-our}, here
we perform
a
rigorous
%simultaneous
fit to precision BCDMS, H1 and ZEUS experimental data\cite{BCDMS-F2,H1-F2-1,H1-F2-2,H1-F2-3,H1+ZEUS-F2,ZEUS-F2} on the proton %deep inelastic
structure function $F_2(x, Q^2)$
to determine necessary
phenomenological parameters
involved into the formulas\cite{PDFs-our}.
%of analytical solutions of the DGLAP equations,
%of DGLAP equations,
The data\cite{BCDMS-F2,H1-F2-1,H1-F2-2,H1-F2-3,H1+ZEUS-F2,ZEUS-F2} cover extremely wide kinematical region,
%taken %in a wide kinematical region, %of $x$ and $Q^2$ namely,
$2 \cdot 10^{-5} < x < 0.75$ and $1.2 < Q^2 < 30000$~GeV$^2$.
Then we derive
analytical expressions for the proton TMDs with
%TMD parton distributions in a proton
%with taking into account
different treatment of kinematical constraint.
This constraint is used in the KMR/WMR prescription and corresponds to the angular or strong ordering condition for parton emissions
at the last evolution step.
%Moreover, we discuss
%relations between the differential and integral formulation of the KMR procedure.
Finally, we test the obtained TMDs
with beauty production
in $pp$ collisions at the LHC. %energies $\sqrt s = 2.76$, $5.02$ and $7$~TeV
Here we rely on our previous investigations\cite{bb-kt-our1, bb-kt-our2}.
We calculate total and differential
cross sections for $b$-jet and $b\bar b$-dijet events
and compare our predictions with available CMS\cite{cjet-CMS,bjet-CMS-7} and ATLAS\cite{bjet-ATLAS-7} experimental data
 taken at $\sqrt s = 7$~TeV.

Our present study is the continuation of previous investigations\cite{TMDs-KMR-our-1,TMDs-KMR-our-2,nTMDs-KMR-our},
where small-$x$ asymptotics
for the TMD parton densities in a proton and nuclei
were considered. %\footnote{Nuclear TMD parton distributions (nTMDs) have been derived recently in the framework of PB approach\cite{nTMDs-PB}.}.
Moreover, it significantly improves our preliminary results\cite{PDFs-our},
where some of phenomenological parameters of the analytical solution of the DGLAP evolution equations
have been determined rather roughly.
The main advantage of the developed approach is related with a quite compact
analytical formulas for the TMD parton densities in a proton,
which could be easily applied in further phenomenological studies
and also could be extended to the nuclear distributions.

The outline of our paper is following. In Section~2 we describe our theoretical input.
Here we list small-$x$ and large-$x$ asymptotics which are used to construct the PDFs
in a whole kinematical region.
Section~3 is devoted to determination of the TMD parton densities in the KMR/WMR framework, where all
calculations are explained in detail.
Section~4 presents our numerical results and discussions. Section~5 contains our
conclusions.

\section{The model} \indent

Here we shortly review small-$x$ and large-$x$ asymptotics of the
conventional PDFs presented previously\cite{PDFs-our}.
For the reader's convenience, we also list the basic formulas of the LO KMR/WMR approach itself.

\subsection{Proton PDFs and structure function $F_2(x,Q^2)$} \indent

We start from some basic formulas.
As it is known, the proton structure function $F_2(x,Q^2)$ at the LO of perturbative QCD expansion
can be presented in the simple form (see, for example, \cite{F2-KK-NLO,F2-KKS-NNLO})
\begin{gather}
  F_2(x,Q^2) =F_2^{\rm LT}(x,Q^2)\,\left(1+\frac{h(x)}{Q^2}\right),
  \label{eq1-F2}
\end{gather}
\noindent
with the leading twist (LT) part
%As it is well known, the proton structure function $F_2(x,Q^2)$ at LO of perturbative QCD expansion
%can be presented in the simple form
\begin{gather}
  F_2^{\rm LT}(x,Q^2) = \sum_{i}^{N_f} e_i^2 \left[f_{q_i}(x,Q^2) + f_{\bar{q}_i}(x,Q^2) \right],
\label{eq-F2}
\end{gather}
\noindent
where $e_i$ is the fractional electric charge of quark $q_i$ and
$f_{q_i}(x,Q^2)$ and $f_{\bar q_i}(x,Q^2)$ are the quark and antiquark
densities in a proton (multiplied by $x$), respectively.
The magnitude $h(x)$ of the twist-four part is poorly known theoretically and usualy can be represented in the form
\be
h(x)=h_0+\frac{h_1}{1-x}, \quad \mbox{or} \quad h(x)=\overline{h}x^{-\delta}+ h_0+\frac{h_1}{1-x}
\label{eq1-F2pQCD}
\ee
with $\delta \sim 0.18\div 0.2$\cite{F2-KK-NLO} and $h_0$, $h_1$ and $\bar h$
being free parameters.

Everywhere below we use the %FFNS
fixed-flavor-number-sheme\footnote{The FFNS results with $N_f = 4$ can be obtained in the simplest way. Moreover, the results are usually very similar
(see \cite{F2-KKS-NNLO}) to ones obtained in the variable-flavor-number-sheme (VFNS). However, the application of the latter is more complicated.
We plan to perform VFNS studies in out future investigations.} (FFNS)
with $N_f = 4$.
In this scheme, where $b$ and $t$ quarks are separated out,
we have
\begin{gather}
  F_2^{\rm LT}(x,Q^2) = \frac{5}{18} \, f_{SI}(x,Q^2) + \frac{1}{6} f_{NS}(x,Q^2).
\label{eq2-F2-NF4}
\end{gather}
\noindent
In the formulas above the singlet part $f_{SI}(x,\mu^2)$ contains the valence ($V$) and sea ($S$) quark parts:
\begin{gather}
  f_{V}(x,\mu^2) = f_u^V(x,\mu^2) + f_d^V(x,\mu^2), \quad f_{S}(x,\mu^2) = \sum_{i=1}^{N_f} \left[f_{q_i}^S(x,\mu^2) + f_{\bar{q}_i}^S(x,\mu^2) \right], \nonumber \\
  f_{SI}(x,\mu^2) = \sum_{i=1}^{N_f} \left[f_{q_i}(x,\mu^2) + f_{\bar{q}_i}(x,\mu^2) \right] = f_{V}(x,\mu^2) + f_{S}(x,\mu^2).
\label{eq-FSI}
\end{gather}
\noindent
The nonsinget part $f_{NS}(x,\mu^2)$ contains difference between the up and down quarks:
\begin{gather}
f_{NS}(x,\mu^2) = \sum_{q=u,\,c} \left[f_{q}(x,\mu^2) + f_{\bar{q}}(x,\mu^2) \right] - \sum_{q=d,\,s} \left[f_{q}(x,\mu^2) + f_{\bar{q}}(x,\mu^2) \right].
\label{eq-FNS-NF4}
%\end{align}
\end{gather}
\noindent
All these formulas above are very useful in accurate determination of
phenomenological parameters of proton PDFs derived in our previous paper\cite{PDFs-our}.
Note that approach\cite{PDFs-our} follows the idea\cite{PDFs-our-previous} and consists
of two basic steps. First of them is the analytical calculation of the asymptotics of solutions of the DGLAP
equations for the parton densities at low and large values of Bjorken variable $x$. Second,
to obtain the analytical expressions for PDFs over the full $x$ range,
one can combine these solutions and then interpolate between them (see also\cite{PDFs-early-approach}).

\subsection{Non-singlet and valence quark parts} \indent

The non-singlet and valence parts of quark distribution functions can be represented in the following form\cite{PDFs-our}:
\begin{gather}
  f_i(x,\mu^2) = \left[A_i(s)x^{\lambda_i}(1 -x) + \frac{B_i(s)\, x}{\Gamma(1+\nu_i(s))} + D_i(s)x (1 -x) \right] (1-x)^{\nu_i(s)},
\label{eq1-NSV}
\end{gather}
\noindent
where $i = NS$ or $V$, $s = \ln \left[\alpha_s(Q_0^2)/\alpha_s(\mu^2)\right]$ and
\begin{gather}
  A_{i}(s)=A_{i}(0) e^{-d(n_i)s}, \quad B_{i}(s) = B_{i}(0) e^{-p s}, \quad \nu_{i}(s)=\nu_{i}(0)+r s, \nonumber \\
  r = \frac{16}{3\beta_0}, \quad p=r\left(\gamma_{\rm E}+\hat{c}\right), \quad \hat{c}=-\frac{3}{4}, \quad d(n)=\frac{\gamma_{NS}(n)}{2\beta_0}, \quad n_i = 1 - \lambda_i.
\label{eq2-NSV}
\end{gather}
\noindent
Here $\Gamma(z)$ is the Riemann's $\Gamma$-function, $\gamma_{\rm E} \simeq 0.5772$ is the Euler's constant,
$\beta_0 = 11 - (2/3)N_f$ is the LO QCD $\beta$-function and $\gamma_{NS}(n)$ is the LO non-singlet anomalous
dimension.
From the Regge calculus, $\lambda_{NS} \simeq \lambda_V \simeq 0.3 - 0.5$.
For the simplicity, here we set $\lambda_{NS} = \lambda_V = 0.5$.
Everywhere below, we apply "frozen" treatment of the QCD coupling
in the infrared region (see, for
example,\cite{frozen-aQCD,gDAS-PDFs-smallx} and references therein), where $\alpha_s(\mu^2) \to \alpha_s(\mu^2 + M_\rho^2)$
with $M_\rho \sim 1$~GeV, that immediately leads to $s > 0$.
Such treatment results in a good description of the
data on proton structure function $F_2(x,Q^2)$ at low $Q^2$\cite{frozen-aQCD}.

As it was mentioned above, the expression (\ref{eq1-NSV}) is constructed as the combination of the small-$x$
part proportional to $A_i(s)$, large-$x$ asymptotic proportional to $B_i(s)$ and additional
term proportional to $D_i(s)$. The latter is subasymptotics in both these regions
and fixed by the Gross-Llewellyn-Smith and Gottfreed sum rules\cite{PDFs-our}:
\begin{gather}
  D_i(s) = \left(2 + \nu_i(s) \right) \left[N_i - A_i(s) {\Gamma(\lambda_i) \Gamma(2 + \nu_i(s)) \over \Gamma(\lambda_i + 2 + \nu_i(s))} - {B_i(s) \over \Gamma(2 + \nu_i(s))} \right],
\label{eq3-NSV}
\end{gather}
\noindent
where $N_V = 3$ and
\begin{gather}
  N_{NS} = 1 + 2 \int\limits_0^1 {dx\over x} \left[ f_{\bar u}(x,\mu^2) + f_{\bar c}(x,\mu^2) - f_{\bar d}(x,\mu^2) - f_{\bar s}(x,\mu^2)\right] \equiv  I_G(\mu^2).
\end{gather}
\noindent
%for $N_f = 4$.
%$N_{NS} = I_G(\mu^2)$ for flavor-symmetric sea.
The $\mu^2$-dependence of
$I_G(\mu^2)$ is very weak\cite{IG-1} and, therefore, can be safely neglected in our LO analysis.
So, everywhere below we use $I_G(\mu^2) \equiv I_G = 0.705 \pm 0.078$\cite{NMC}.
%and then $N_{NS} = 3 + 2(I_G - 1) = 2.410 \pm 0.156$ for $N_f = 3$.
%Moreover, one can set $A_{NS}(0) = A_V(0) I_G/9$ and $B_{NS}(0) = B_V(0) I_G/9$\cite{?}.

Free parameters $A_i(0)$, $B_i(0)$ and $\nu_i(0)$
involved in~(\ref{eq1-NSV}) and (\ref{eq2-NSV}) %and $Q_0^2$
%are not predicted by the theory and
can be determined, for example, from the direct
comparison with appropriate numerical solutions\cite{CT14}
of the DGLAP equations or from experimental data (see below).

\subsection{Singlet quark and gluon parts} \indent

The sea and gluon densities in a proton can be represented as combinations of "$\pm$" terms
in the following form\cite{PDFs-our} (see also\cite{PDFs-our-previous-2}):
\begin{gather}
  f_i(x,\mu^2) = f_i^+(x,\mu^2) + f_i^-(x,\mu^2),
\label{eq1-Sg}
\end{gather}
\noindent
where $i = SI$ or $g$ and
\begin{gather}
  f_{SI}^-(x,\mu^2) = \tilde{f}_{SI}^{-}(x,\mu^2)(1 - x)^{\nu^{-}(s)}, \quad f_{SI}^{+}(x,\mu^2) = \tilde{f}_{SI}^{+}(x,\mu^2)(1 - x)^{\nu^{+}(s) + 1}, \nonumber \\
  f_{g}^-(x,\mu^2) = \tilde{f}_{g}^-(x,\mu^2)(1 - x)^{\nu^{-}(s)+1}, \quad f_{g}^+(x,\mu^2) = \tilde{f}_{g}^+(x,\mu^2)(1 - x)^{\nu^{+}(s)},
\label{tPDFs}
\end{gather}
with
\begin{gather}
  \tilde{f}_{SI}^+(x,\mu^2) = {N_f \over 9} \left[A_{g} + {4\over 9}A_q\right] \rho I_1(\sigma) e^{- \bar d^{+} s} (1-x)^{m_{q}^+} + D^+(s) \sqrt x (1 - x)^{n^+} - \nonumber \\
  - {K^+ \over \Gamma(2 + \nu^+(s))} \times {B^+(s)x \over \hat c - \ln(1 - x) + \Psi(2 + \nu^+(s))},
\label{eq-fSp}
\end{gather}
\begin{gather}
  \tilde{f}_{SI}^-(x,\mu^2) = A_{q}e^{- d^{-} s} (1-x)^{m_{q}^-} + {B^-(s)x \over \Gamma(1 + \nu^-(s))} + D^{-}(s) \sqrt x (1 -x)^{n^-},
\label{eq-fSm}
\end{gather}
\begin{gather}
  \tilde{f}_{g}^+(x,\mu^2) = \left(A_{g} + {4\over 9}A_q\right) I_0(\sigma) e^{- \bar d^{+} s}(1 - x)^{m_{g}^+} + {B^+(s) x\over \Gamma(1 + \nu^+(s))},
\label{eq-fgp}
\end{gather}
\begin{gather}
  \tilde{f}_{g}^-(x,\mu^2) = - {4\over 9} A_{q} e^{- d^{-} s} (1 - x)^{m_g^-} + \\ \nonumber
  + {K^- \over \Gamma(2 + \nu^-(s))} \times {B^-(s)x \over \hat c - \ln(1 - x) + \Psi(2 + \nu^-(s))}.
\label{eq-fgmtld}
\end{gather}
\noindent
Here $\Psi(z)$ is the Riemann's $\Psi$-function, $I_0(z)$ and $I_1(z)$ are the modified Bessel functions and
\begin{gather}
  \nu^\pm(s) = \nu^\pm(0) + r^\pm s, \quad B^\pm(s) = B^\pm(0) e^{-p^\pm s}, \quad p^\pm = r^\pm (\gamma_{\rm E} + \hat c^\pm), \nonumber \\
  r^+ = {12\over \beta_0}, \quad r^- = {16\over 3 \beta_0}, \quad \hat c^+ = - {\beta_0 \over 12}, \quad \hat c^- = - {3\over 4}, \quad K^+ = {3N_f \over 10}, \quad K^- = {2\over 5}, \nonumber \\
  \rho = {\sigma \over 2 \ln(1/x)}, \quad \sigma = 2 \sqrt{|\hat d^+|s \ln{1\over x}}, \\ \nonumber
  \hat d^+ = - {12\over \beta_0}, \quad \bar d^+ = 1 + {20 N_f\over 27\beta_0}, \quad d^- = {16 N_f \over 27 \beta_0}
\label{eq-parameters}
\end{gather}
\noindent
with $n^+ \sim n^-$ being free parameters.
The small-$x$ PDFs asymptotics seen in (\ref{eq-fSp}) --- (\ref{eq-fgmtld})
were obtained at LO\cite{gDAS-PDFs-smallx,gDAS1,gDAS2,gDAS-PDFs-smallx-NLO} in the so-called generalized
doubled asymptotic scaling (DAS) approximation\cite{gDAS1,gDAS2,gDAS3}.
In this approximation, flat initial conditions for proton PDFs, $f_g(x,Q_0^2) = A_g$ and $f_S(x,Q_0^2) = A_q$, can be
used\footnote{The NLO analysis has been done\cite{gDAS1, gDAS2}.}.
In the numerical analysis below we set $m_q^- = m_g^+ = 2$,  $m_q^+ = m_g^- = 8$.
%with $N_f = 4$ and $m_q^- = m_g^+ = 1$,  $m_q^+ = m_g^- = 5$ with $N_f = 3$, respectively.
In this case, small-$x$ asymptotics above are suppressed at large $x$ compared to
subasymptotic terms proportional to $D^\pm(s)$. Moreover, these small-$x$ asymptotics contain the same
powers of $(1 - x)$ factor for both quarks and gluons.

The expressions for $D^\pm(s)$ could be derived from the momentum conservation law\cite{PDFs-our}:
\begin{gather}
D^+(s) = - \frac{\Gamma(7/2 + n^+ + \nu^+(s))}{\Gamma(3/2)\Gamma(2+n^+ + \nu^+(s))} \left[G_{SI}^+(s) + \bar G_g^+(s)\right], \nonumber \\
D^-(s) =  \frac{\Gamma(5/2 + n^- + \nu^-(s))}{\Gamma(3/2)\Gamma(1+n^- + \nu^-(s))} \left[1- G_g^-(s) - \bar G_{SI}^-(s)\right],
\label{eq1-Dpm}
\end{gather}
\noindent
where
\begin{gather}
  G_{SI}^+(s) = {N_f \over 9}\left[A_g + {4\over 9} A_q\right] \Phi_1(1 + m_q^+ + \nu^+(s)) e^{-\bar d^+ s} - \nonumber \\
 - {K^+B^+(s) \over \Gamma(4 + \nu^+(s)) (\Psi(4 + \nu^+(s)) + \hat c)}, \\
  \bar G_g^+(s) = \left[A_g + {4\over 9} A_q\right] \Phi_0(m_g^+ + \nu^+(s)) e^{-\bar d^+ s} + {B^+(s) \over \Gamma(3 + \nu^+(s))}, \\
%  G_V(s) = {1\over 3 + \nu_V(s)} \left[ N_V - (1 - \lambda_V)A_V(s){\Gamma(\lambda_V) \Gamma(3 + \nu_V(s))\over \Gamma(\lambda_V + 3 + \nu_V(s))} + {B_V(s) \over \Gamma(3 + \nu_V(s))} \right], \\
  G_g^-(s) = - {4 A_q\over 9(2 + \nu^-(s) + m_g^-)} e^{-d^-s} + {K^-B^-(s) \over \Gamma(4 + \nu^-(s)) (\Psi(4 + \nu^-(s)) + \hat c)},\\
  \bar C_{SI}^-(s) = {A_q \over 1 + \nu^-(s) + m_q^-} e^{-d^-s} + {B^-(s) \over \Gamma(3 + \nu^-(s))},
\label{eq2-Dpm}
\end{gather}
\noindent
and
\begin{gather}
  \Phi_0(\nu) = \int\limits_0^1 dx I_0(\sigma) (1 - x)^{\nu(s)}, \quad \Phi_1(\nu) = \int\limits_0^1 dx \rho I_1(\sigma) (1 - x)^{\nu(s)}.
\end{gather}
\noindent
%From the quark counting rules the relation $\nu^+(s) = \nu^-(s) + 1$ can be obtained, where $\nu^-(0) \sim 3$.
At large $\nu$ one can use the following approximation (see\cite{PDFs-our} for more details):
\begin{gather}
  \Phi_0(\nu) \simeq {1 \over 1 + {\nu}} I_0\left(2 \sqrt{|\hat d_+| s \left(\ln(1 + \nu) + \gamma_{\rm E}\right) }\right), \\
  \Phi_1(\nu) \simeq {1 \over 1 + {\nu}} \left(\sqrt{|\hat d_+| s \over \ln(1 + \nu) + \gamma_{\rm E} }\right)
  I_1\left(2 \sqrt{|\hat d_+| s \left(\ln(1 + \nu) + \gamma_{\rm E}\right) }\right).
\end{gather}
\noindent
The exact values of all free parameters $A_g$, $A_q$, $B^\pm(0)$, $\nu^\pm(0)$ and $n^\pm$ could be taken from the fit to
experimental data.

\subsection{LO KMR/WMR approach} \indent

As it was mentioned above, the KMR/WMR approach
is a formalism to construct the TMD parton distributions from well-known
ordinary PDFs. The key assumption is that
the transverse momentum dependence of the parton densities enters only at the last
evolution step. This procedure is believed to take into account effectively the major
part of next-to-leading logarithmic (NLL) terms $\alpha_s (\alpha_s \ln \mu^2/\Lambda_{\rm QCD}^2)^{n-1}$
compared to the leading logarithmic approximation (LLA), where terms proportional to
$\alpha_s^n \ln^n \mu^2/\Lambda_{\rm QCD}^2$ are taken into
account.

There are known differential $(d)$ and integral ($i$) definitions of the KMR/WMR
prescription. According to these definitions, the TMD parton densities in a proton
can be calculated as\cite{KMR-LO}
\begin{gather}
  f_a^{(d)}(x, {\mathbf k}_T^2, \mu^2) = {\partial \over \partial \ln {\mathbf k}_T^2} \left[ T_a({\mathbf k}_T^2, \mu^2) f_a(x,{\mathbf k}_T^2) \right] \label{eq-KMR-diff}, \\
  f_a^{(i)}(x, {\mathbf k}_T^2, \mu^2) = 2 a_s({\mathbf k}_T^2) T_a({\mathbf k}_T^2, \mu^2) \sum_b \int\limits_x^{1 - \Delta} dz\, P_{ab}(z) f_b\left({x\over z}, {\mathbf k}_T^2\right),
\label{eq-KMR-int}
\end{gather}
\noindent
where $a = q$ or $g$, $a_s(\mu^2) = \alpha_s(\mu^2)/4\pi$, $T_a({\mathbf k}_T^2, \mu^2)$ are the Sudakov form factors and $f_a(x,\mu^2) = x D_a(x,\mu^2)$ are the standard PDFs obeying the DGLAP equations:
\begin{gather}
  \frac{\partial D_a(x,\mu^2)}{\partial \ln \mu^2} = 2a_s(\mu^2) \sum_{b} \int\limits^{1 - \Delta}_x \frac{dz}{z} P_{ab}(z,\mu^2) D_{b}\left(\frac{x}{z},\mu^2 \right) - \nonumber \\
  - D_a(x,\mu^2) \sum_{b} \int\limits^{1 - \Delta}_0 dz z P_{ba}(z,\mu^2).
\label{DGLAPd}
\end{gather}
\noindent
Here $\Delta$ is some cutoff parameter (see below) and
$P_{ab}(z)$ are the unregulated LO DGLAP splitting functions:
\begin{gather}
 P_{qq}(z) = C_F \left[\frac{1+z^2}{(1-z)_+} + \frac{3}{2} \delta (1-z) \right], \quad P_{qg}(z) = N_f \Bigl[z^2 + (1-z)^2\Bigr], \nonumber \\
 P_{gq}(z) = C_F \left[\frac{1+(1-z)^2}{z}\right], \nonumber \\
 P_{gg}(z) = 2C_A \left[\frac{z}{(1-z)_+} + \frac{1-z}{z} + z(1-z) + \frac{\beta_0}{12}  \delta (1-z) \right].
\label{splitingLO}
\end{gather}
\noindent
The quark and gluon Sudakov form factors $T_a({\mathbf k}_T^2, \mu^2)$ give the probability of evolving from a
scale ${\mathbf k}_T^2$ to a scale $\mu^2$ without parton emission:
\begin{gather}
  \ln T_a({\mathbf k}_T^2, \mu^2) = - 2\int\limits_{{\mathbf k}_T^2}^{\mu^2} {d {\mathbf p}_T^2 \over {\mathbf p}_T^2} a_s({\mathbf p}_T^2)
  \sum_b \int\limits_0^{1 - \Delta} dz z P_{ba}(z).
\label{eq-sudakov}
\end{gather}
\noindent
As it is often done, we set $T_a({\mathbf k}_T^2, \mu^2) = 1$ at ${\mathbf k}_T^2 > \mu^2$.

Several important remarks are in order.
First, both integrals appearing in~(\ref{DGLAPd}), %the DGLAP equations
which describe real and virtual parton emissions,
are divergent at $\Delta = 0$ due to the singular
splitting functions $P_{qq}(z)$ and $P_{gg}(z)$ at $z = 1$.
However, these singularities, which are due to soft radiation, cancel out when the two terms
are combined through the "$+$" prescription.
In contrast,
if the upper integration limits in~(\ref{DGLAPd}) and~(\ref{eq-sudakov}) are restricted by some $\Delta > 0$,
then $\delta$-functions in~(\ref{splitingLO}) give no contributions. Moreover, the "$+$" prescription is not needed: $(1-z)^{-1}_+ \to (1-z)^{-1}$.
In this case, the positive real emissions are separated
from negative virtual ones, that could results in some
ambiguities in the KMR/WMR formalism (see, for example, discussions\cite{KMR-discussion-1, KMR-discussion-2, KMR-discussion-3, KMR-discussion-4}).
So, the differential and integral KMR/WMR definitions are not equivalent
if ordinary PDFs are used in~(\ref{eq-KMR-diff}) and (\ref{eq-KMR-int}).
In particular, the difference between two these definitions occur at large transverse momenta.
It was shown\cite{KMR-discussion-1} that (\ref{eq-KMR-diff}) and (\ref{eq-KMR-int}) become
mathematically equivalent if special cutoff $\Delta$-dependent PDFs are employed
instead of standard PDFs (as obtained from the DGLAP equations at $\Delta = 0$
and global fits to data).
Nevertheless, the TMD parton densities derived from the integral definition~(\ref{eq-KMR-int})
are practically the same, regardless on the choice of the ordinary or cutoff dependent PDFs in the calculations\cite{KMR-discussion-1}.
Thus, this definition seems to be more preferable in the phenomenological applications.
Concerning the cutoff parameter $\Delta$, usually it has one of two basic
physically motivated forms:
\begin{gather}
  \Delta_1 = {|{\mathbf k}_T| \over \mu}, \quad \Delta_2 = {|{\mathbf k}_T| \over |{\mathbf k}_T| + \mu},
\label{eq-deltas}
\end{gather}
\noindent
that reflects the strong ordering (SO) or angular ordering (AO) conditions for parton emissions
at the last evolution step\footnote{The strong ordering in transverse momentum
within the DGLAP equations automatically ensures angular ordering.}
to regulate the soft gluon singularities (see\cite{KMR-LO} for
more information).
It is important that parton distributions $f_a(x, {\mathbf k}_T^2, \mu^2)$ extend
into the ${\mathbf k}_T^2 > \mu^2$ region in the case
of angular ordering and vanish at large ${\mathbf k}_T^2$
in the case of strong ordering condition.

Another point is related to the treatment of low ${\mathbf k}_T^2$ region. In fact, both
the differential and integral definitions of the KMR/WMR approach are only correct at ${\mathbf k}_T^2 > \mu_0^2$,
where $\mu_0^2 \sim 1$~GeV$^2$ is the minimum scale where perturbative QCD is still applicable.
At small ${\mathbf k}_T^2$ special model assumptions are necessary.
Usually, such assumptions are connected with the normalization condition
\begin{gather}
  f_a(x,\mu^2) = \int\limits_0^{\mu^2_{\rm max}} f_a(x,{\mathbf k}_T^2, \mu^2) d {\mathbf k}_T^2,
 \label{eq-norm}
\end{gather}
\noindent
where $\mu^2_{\max}$ is taken to be equal to $\mu^2$ (see, for example,\cite{KMR-LO, KMR-discussion-1})
or even infinity\cite{KMR-discussion-4}.
The rather arbitrary shape of $f_a(x,{\mathbf k}_T^2, \mu^2)$ at low ${\mathbf k}_T^2 < \mu_0^2$ is often chosen, such as
flat or Gaussian-like.
However, in our consideration, where "frozen" treatment of the QCD coupling in the infrared region
is applied, the TMDs are well defined in the whole ${\mathbf k}_T^2$ range.
Thus, we do not use the normalization condition~(\ref{eq-norm}). Moreover,
in our opinion, any relations between the TMDs and collinear PDFs
can only be approximate since these quantities are essentially different objects
where essentially different large logarithmic terms appearing in the perturbative QCD expansion are resummed.

\section{Analytic results} \indent

The analytical expressions for conventional PDFs collected above
could be used as an input for the KMR/WMR procedure,
giving us the possibility to derive the analytical expressions for the TMD parton densities in a proton.

\subsection{Differential definition} \noindent

Now we calculate the TMD parton densities according to~(\ref{eq-KMR-diff})
without derivatives.
Derivation of the $T_a(\mu^2,k^2)$ and conventional PDFs are as follows
\begin{gather}
  \frac{\partial T_a({\mathbf k}_T^2, \mu^2)}{\partial \ln {\mathbf k}_T^2} = d_a \beta_0 a_s({\mathbf k}_T^2) R_a(\Delta) \label{DefTa}, \\
  \frac{\partial f_a(x,{\mathbf k}_T^2)}{\partial \ln {\mathbf k}_T^2} \approx -\beta_0 a_s({\mathbf k}_T^2) \times \\
  \times \left[  \left(\frac{\hat{d}^+}{\rho_a}  + \overline d^{+}-r^+ \ln(1-x)\right) f_a^+(x, {\mathbf k}_T^2) + \left(d^{-} -r^- \ln(1-x)\right) f_a^-(x, {\mathbf k}_T^2) \right],
\label{Deffa}
\end{gather}
\noindent
where
\begin{gather}
  R_q(\Delta) = \ln\left(\frac{1}{\Delta}\right) - \frac{3}{4} (1-\Delta)^2, \nonumber \\
  R_g(\Delta) = \ln\left(\frac{1}{\Delta}\right) - \left(1-\frac{\varphi}{4}\right) (1-\Delta)^2 + \frac{1-\varphi}{12} (1-\Delta)^3 (1+3\Delta), \nonumber \\
  d_a = {4 C_a\over \beta_0}, \quad  \hat{d}^+= -|\hat{d}^+|, \quad \varphi = {N_f/C_A}, \quad C_q = C_F, \quad C_g = C_A, \nonumber \\
  \frac{1}{\rho_g} = {1 \over \rho} {I_1(\sigma) \over I_0(\sigma)}, \quad \frac{1}{\rho_q} = {1\over \rho} {I_0(\sigma) \over I_1(\sigma)}.
\label{rho.a}
\end{gather}
\noindent
We note that the result (\ref{Deffa}) is a simple extension of the one\cite{KMR-DAS-1} to the whole $x$ range.
The extension contains the additional terms $\sim \ln(1-x)$.
Nevertheless, we checked that~(\ref{Deffa}) approximates with good accuracy the $\ln {\mathbf k}_T^2$-derivative
of the exact results (\ref{eq1-Sg}) --- (\ref{eq-fgp}). %for $f_a^{\pm}(x,{\mathbf k}_T^2)$.}
So, the final formula for the TMD parton densities reads
\begin{gather}
f^{(d)}_a(x,{\mathbf k}_T^2,\mu^2) \approx \beta_0 a_s({\mathbf k}_T^2) T_a({\mathbf k}_T^2, \mu^2) \times \Biggl(d_a R_a(\Delta) f_a(x,{\mathbf k}_T^2) -\nonumber \\
 - \left[\left(\frac{\hat{d}_+}{\rho_a} + \overline d_{+}-r^+ \ln(1-x)\right) f_a^+(x,{\mathbf k}_T^2)
  + \left(d^{-} -r^- \ln(1-x)\right) f_a^-(x,{\mathbf k}_T^2) \right] \Biggr).
\label{uPDF}
\end{gather}
\noindent
The expressions for conventional PDFs are listed in the Sections~2.2 and 2.3.
Since the value of $\hat{d}_+$ is negative and factor $\hat{d}_+/\rho_a$ is large at small $x$, the TMD parton
densities $f^{(d)}_a(x,{\mathbf k}_T^2,\mu^2)$ are
positive. With increasing $x$, the results
for the latter can be negative.% that, in particular, demonstrates an inapplicability of(\ref{8.02}) for PDFs.

%So, we see that the expression (\ref{uPDF}) for TMDs in the differential formulation of KMR approach is rather complicated. Moreover, it is an approximation
%and more exact results are essetially more cumbersome (see \cite{PDFs-our}).

\subsection{Integral definition} \noindent

The corresponding results for TMDs in the integral formulation~(\ref{eq-KMR-int}) are
a bit more complicated. So, as it was done earlier\cite{PDFs-our}, here we limit ourselves by an appoximation.
Similar to~(\ref{eq1-Sg}), we represent sea and gluon densities as
combinations of "$\pm$" terms:
\begin{gather}
  f^{(i)}_a(x,{\mathbf k}_T^2,\mu^2) = f^{(i),+}_a(x,{\mathbf k}_T^2,\mu^2) + f^{(i),-}_a(x,{\mathbf k}_T^2,\mu^2),
\label{eq1-Sg_kt}
\end{gather}
\noindent
where $a = SI$ or $g$ and
\begin{gather}
  f_{SI}^{(i),-}(x,{\mathbf k}_T^2,\mu^2) = \tilde{f}_{SI}^{(i),-}(x,{\mathbf k}_T^2,\mu^2) (1-x)^{\nu^{-}(s)}, \nonumber \\
  f_{SI}^{(i),+}(x,{\mathbf k}_T^2,\mu^2) = \tilde{f}_{SI}^{(i),+}(x,{\mathbf k}_T^2,\mu^2) (1-x)^{\nu^{+}(s)+1},\nonumber \\
  f_{g}^{(i),+}(x,{\mathbf k}_T^2,\mu^2) = \tilde{f}_{g}^{(i),+}(x,{\mathbf k}_T^2,\mu^2)  (1-x)^{\nu^{+}(s)}, \nonumber \\
  f_{g}^{(i),-}(x,{\mathbf k}_T^2,\mu^2) = \tilde{f}_{g}^{(i),-}(x,{\mathbf k}_T^2,\mu^2) (1-x)^{\nu^{-}(s)+1}.
\label{tuPDFs}
\end{gather}
\noindent
After some algebra we have
\begin{gather}
 \tilde{f}_{g}^{(i),\pm}(x,{\mathbf k}_T^2,\mu^2)= 4C_g a_s({\mathbf k}_T^2) T_g({\mathbf k}_T^2, \mu^2) \times \left[D^{\pm}_g(x,\Delta)+ \tilde{D}_g^{\pm}(\Delta)\right] \tilde{f}_g^{\pm}\left(\frac{x}{1 - \Delta},{\mathbf k}_T^2\right),
\label{uPDF2.1g}\\
 \tilde{f}_{q}^{(i),\pm}(x{\mathbf k}_T^2,\mu^2)= 4C_q a_s({\mathbf k}_T^2) T_q({\mathbf k}_T^2, \mu^2) \times \nonumber \\
  \left[D^{\pm}_q(x,\Delta)\tilde{f}_q^{\pm}\left(\frac{x}{1 - \Delta},{\mathbf k}_T^2\right) + \overline{D}_q(\Delta) \tilde{f}_g^{\pm}\left(\frac{x}{1 - \Delta},{\mathbf k}_T^2\right)\right],
\label{uPDF2.1q}
\end{gather}
\noindent
where
\begin{gather}
D^{\pm}_g(x,\Delta) = \ln\left(\frac{1-x}{\Delta}\right) - 2(1 - \Delta) + \frac{(1 - \Delta)^2}{2} - \frac{(1 - \Delta)^3}{3} - x S_1(\nu_{\pm}), \nonumber \\
D^{\pm}_q(x,\Delta) = \ln\left(\frac{1-x}{\Delta}\right) - \frac{(1 - \Delta)}{4} (3 - \Delta) - x S_1(\nu_{\pm}), \nonumber \\
\tilde{D}_g^+ = \frac{1}{\overline{\rho}_g} + \frac{4 N_f}{81}, \quad \tilde{D}_g^-= \frac{9(1 - \Delta)}{8}(3 + \Delta), \quad \overline{D}_q = N_f \frac{1 - \Delta}{18} \left(2\Delta^2 - \Delta + 3\right), \nonumber \\
\bar \sigma = \sigma\left(x \to {x\over 1 - \Delta}\right), \quad \bar \rho_a = \rho_a\left(x \to {x\over 1 - \Delta}\right),
\label{uPDF2.3N}
\end{gather}
and $ S_1(\nu)=\Psi(\nu+1)-\Psi(1)$.

We see that expressions (\ref{eq1-Sg_kt}) --- (\ref{uPDF2.3N}) are rather simple extension of the ones\cite{TMDs-KMR-our-1}
to the whole $x$ range. Below we will demonstrate that these
results are in good agreement with the exact numerical
evaluations of $f^{(i)}_a(x,{\mathbf k}_T^2,\mu^2)$ based on~(\ref{eq-KMR-int}) and %$f_a^{\pm}(x,{\mathbf k}_T^2)$ done in
(\ref{eq1-Sg}) --- (\ref{eq-fgp}).

\section{Numerical results} \indent

We are now in position to present our numerical results.
We start from the determination of phenomenological parameters
involved into the derived analytical expressions
for conventional PDFs as well as TMD parton densities in a proton.
Then we test the latter with inclusive heavy jet production at the LHC conditions.

\subsection{Determination of phenomenological parameters} \indent

There are several parameters in the analytical expressions for conventional PDFs
in a proton
which are not predicted by the theory. %and therefore have to be determined from the experimental data.
Some of them, namely, $A_V(0)$, $A_{NS}(0)$, $B_V(0)$, $B_{NS}(0)$,
$\nu_V(0)$ and $\nu_{NS}(0)$, essential in the large-$x$ region, we determine
from the direct comparison with the %parametrizations of
numerical solutions of the DGLAP equations at the starting scale $Q_0^2$.
Here we employ the CT'14 (LO) parametrizations\cite{CT14} obtained
by the CTEQ-TEA group with %$N_f = 3$ and
$N_f = 4$, that fully corresponds our setup.
The remaining parameters %, namely, $A_q$, $A_g$, $B^\pm(0)$ and $\nu^\pm(0)$,
we determine by fitting the precise
data on proton structure function $F_2(x,Q^2)$
taken by the BCDMS\cite{BCDMS-F2}, H1\cite{H1-F2-1,H1-F2-2,H1-F2-3,H1+ZEUS-F2} and ZEUS\cite{H1+ZEUS-F2,ZEUS-F2} Collaborations.
These data %\cite{BCDMS-F2,H1-F2-1,H1-F2-2,H1-F2-3,H1+ZEUS-F2,ZEUS-F2}
cover extremely wide
region of $x$ and $Q^2$, namely, $2 \cdot 10^{-5} < x < 0.75$ and $1.2 < Q^2 < 30000$~GeV$^2$,
thus providing us the possibility to extract
all necessary parameters simultaneously.
Note that we included into the fit procedure low-$Q^2$ data,
where the higher-twist corrections
could play a role. Such corrections are taken into account according to general expressions~(\ref{eq1-F2}) --- (\ref{eq1-F2pQCD}).
Next, we set $n^+ \sim n^-$, $\nu^+(0) \sim \nu^-(0)$ and keep in mind that
the QCD evolution will result in $\nu^+(s) > \nu^-(s)$ for $s > 0$.
Our fit is based on the LO formulas~(\ref{eq2-F2-NF4}) --- (\ref{eq-FNS-NF4}),
where we apply $Q_0^2 = 0.43$~GeV$^2$, as it was done earlier\cite{PDFs-our, PDFs-our-previous-2}.
We use world averaged $\alpha_s(M_Z^2) = 0.1180$\cite{PDG} with
%$\Lambda_{\rm QCD}^{(3)} = 141$~MeV. % and
$\Lambda_{\rm QCD}^{(4)} = 118$~MeV.
%, respectively.

\begin{table} \scriptsize
\label{table1}
\begin{center}
\begin{tabular}{c c c c c c}
\hline
\hline
 & & & & & \\
 $A_q$ & $A_g$ & $A_V(0)$ & $B_V(0)$ & $A_{NS}(0)$ & $B_{NS}(0)$ \\
 & & & & & \\
\hline
 & & & & &\\
 $0.75 \pm 0.02$ & $1.13 \pm 0.01$ & $2.82 \pm 0.01$ & $(1.06 \pm 0.05)\cdot 10^3$ & $0.279 \pm 0.006$ & $(1.19 \pm 0.04)\cdot 10^2$\\
 & & & & &\\
\hline
\hline
 & & & & &\\
 $B^+(0)$ & $B^-(0)$ & $\nu_V(0)$ & $\nu_{NS}(0)$ & $\nu^-(0)$ & $n^-$ \\
 & & & & &\\
\hline
 & & & & &\\
 $(6.2 \pm 0.5)\cdot 10^1$ & $(2.27 \pm 0.10)\cdot 10^2$ & $4.05 \pm 0.02$ & $3.23 \pm 0.01$ & $3.54 \pm 0.02$ & $4.66 \pm 0.05$\\
 & & & & &\\
\hline
\hline
\end{tabular}
\end{center}
\caption{The fitted values of various phenomenological parameters involved in our analytical expressions for
PDFs and TMD parton densities in a proton.}
\label{tbl:parameters4}
\end{table}

\begin{figure}
\begin{center}
\includegraphics[width=2.8cm]{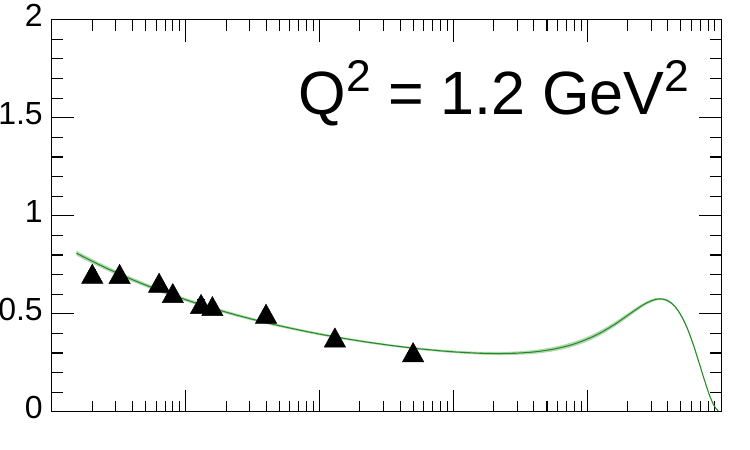}
\hspace*{-0.45cm} \includegraphics[width=2.8cm]{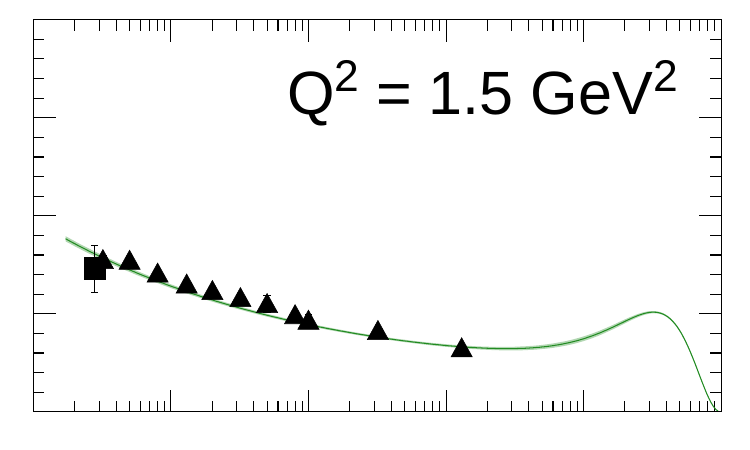}
\hspace*{-0.45cm} \includegraphics[width=2.8cm]{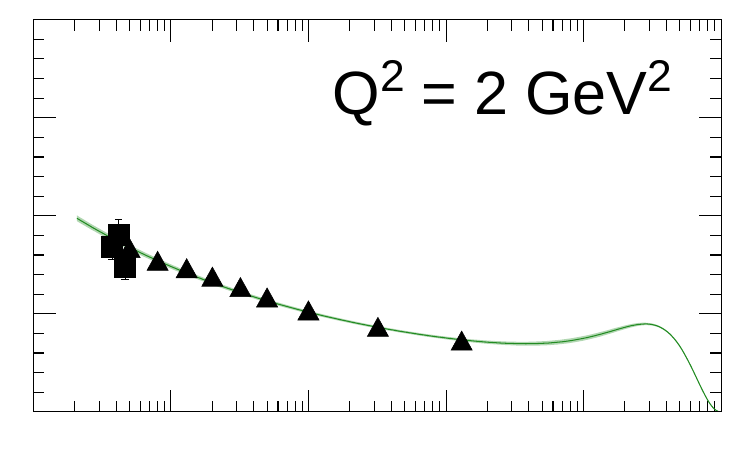}
\hspace*{-0.45cm} \includegraphics[width=2.8cm]{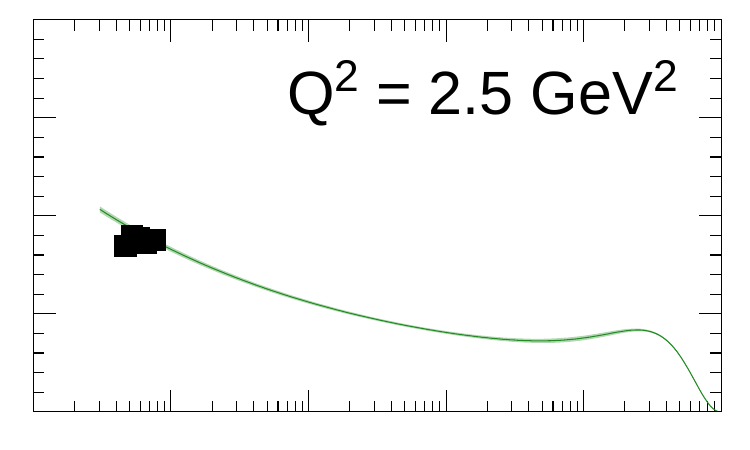}
\hspace*{-0.45cm} \includegraphics[width=2.8cm]{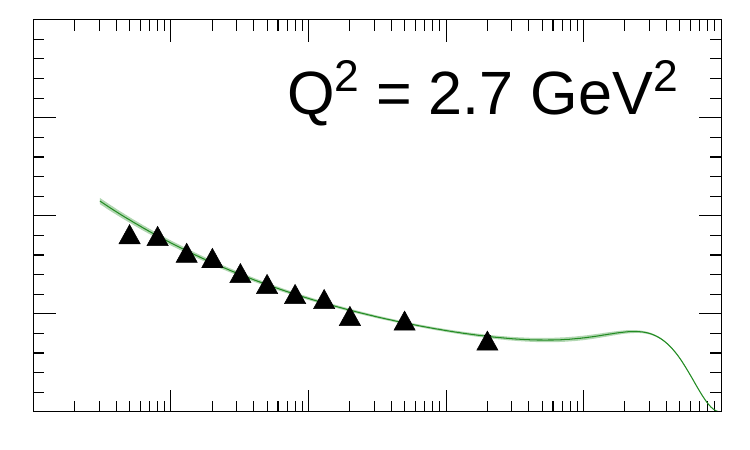}
\hspace*{-0.45cm} \includegraphics[width=2.8cm]{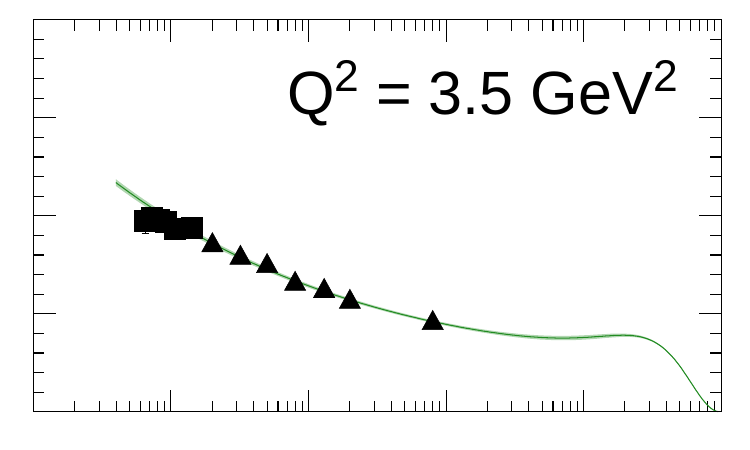}
\includegraphics[width=2.8cm]{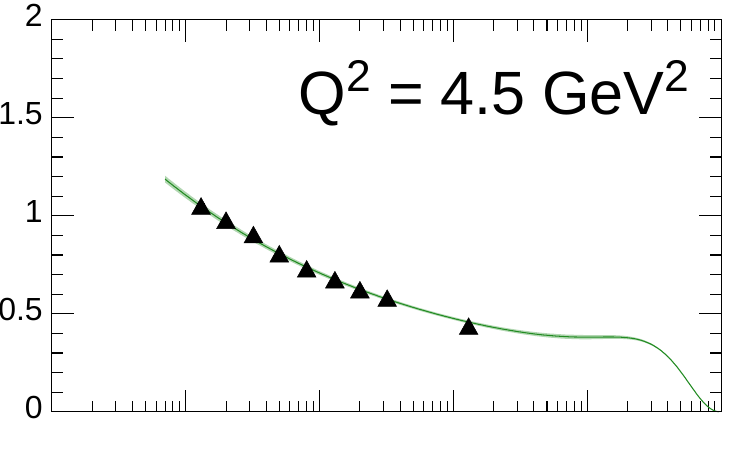}
\hspace*{-0.45cm} \includegraphics[width=2.8cm]{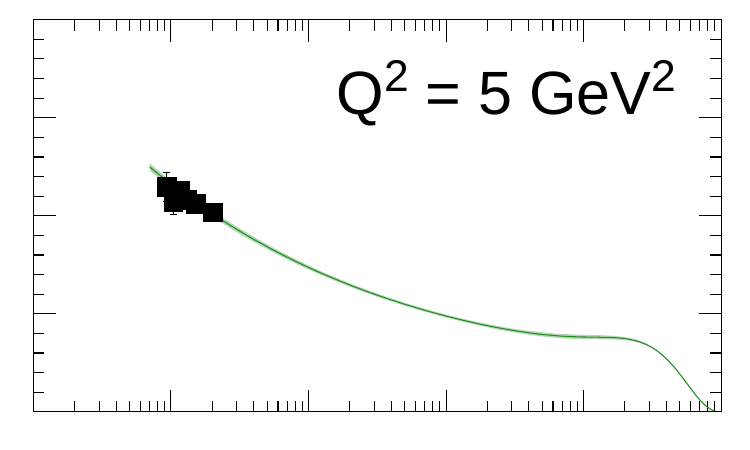}
\hspace*{-0.45cm} \includegraphics[width=2.8cm]{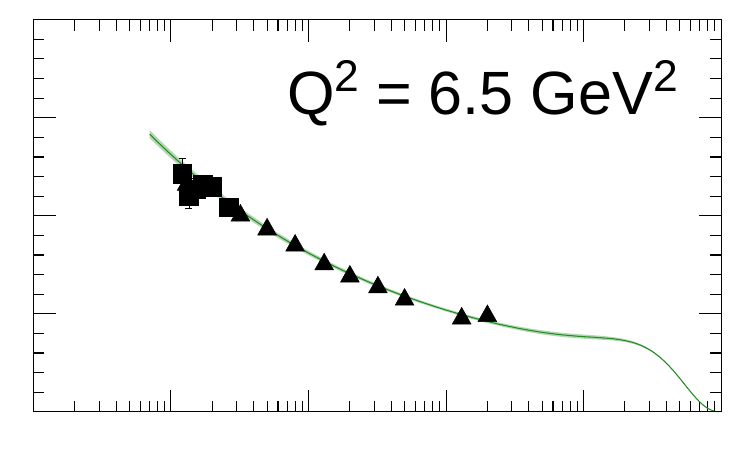}
\hspace*{-0.45cm} \includegraphics[width=2.8cm]{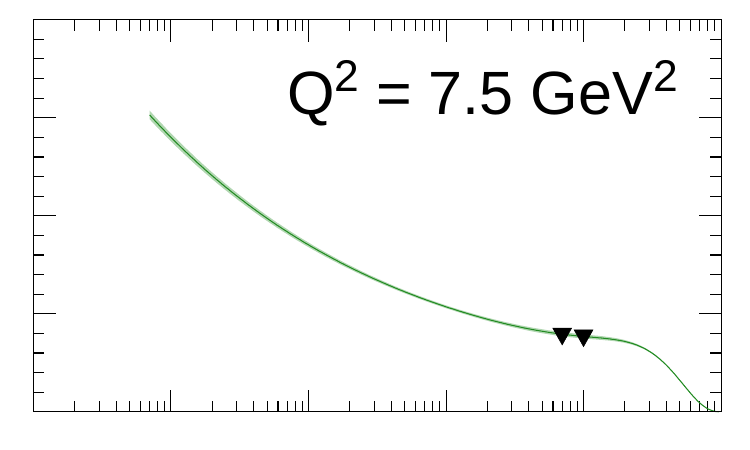}
\hspace*{-0.45cm} \includegraphics[width=2.8cm]{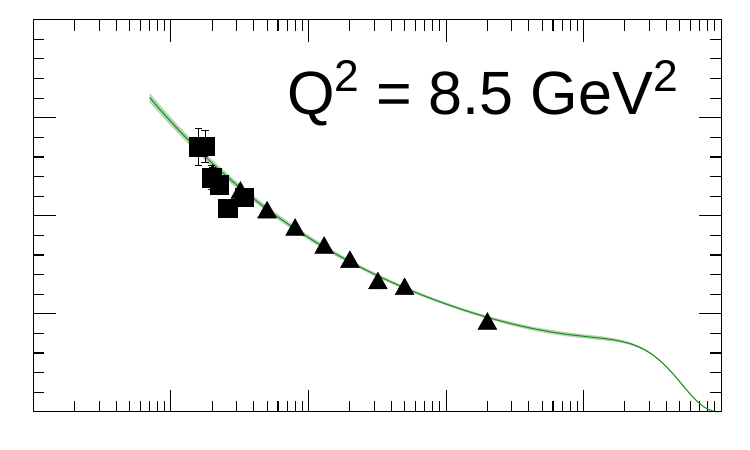}
\hspace*{-0.45cm} \includegraphics[width=2.8cm]{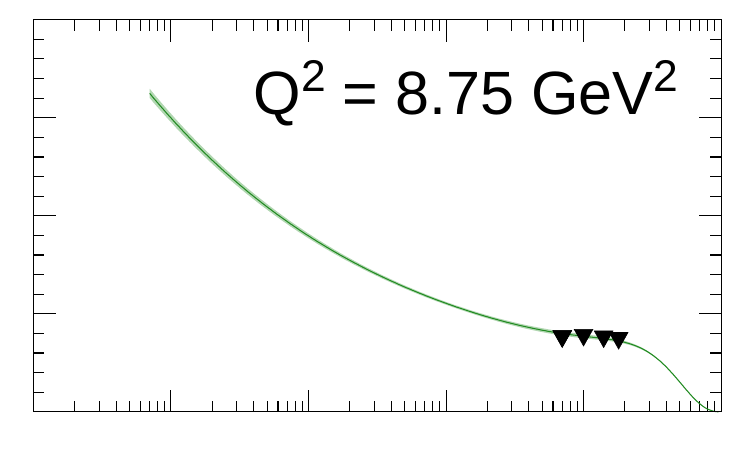}
\includegraphics[width=2.8cm]{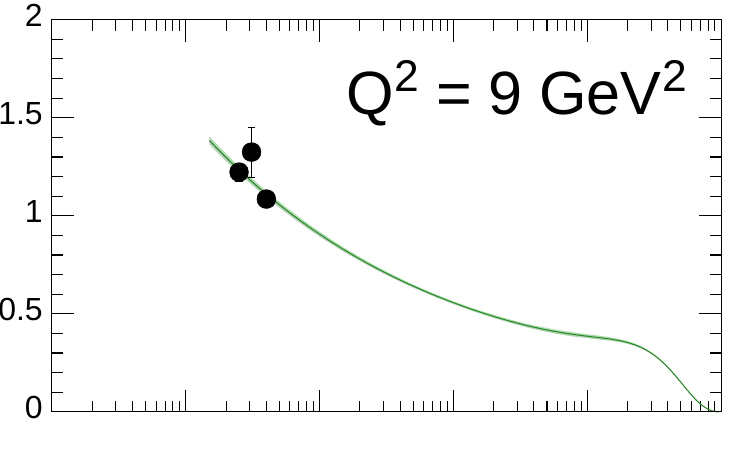}
\hspace*{-0.45cm} \includegraphics[width=2.8cm]{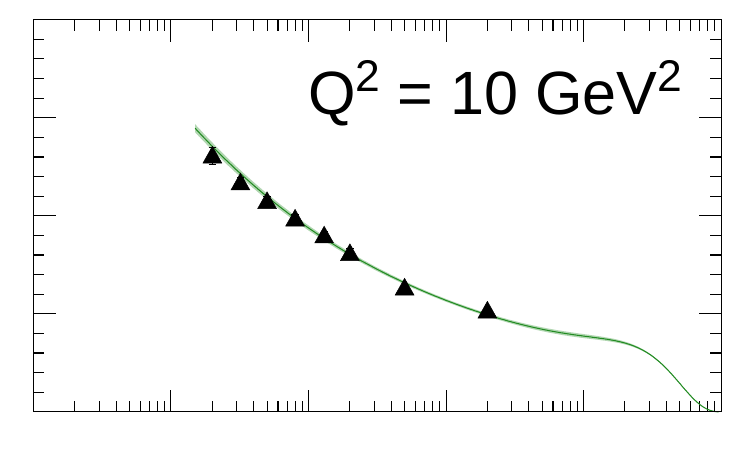}
\hspace*{-0.45cm} \includegraphics[width=2.8cm]{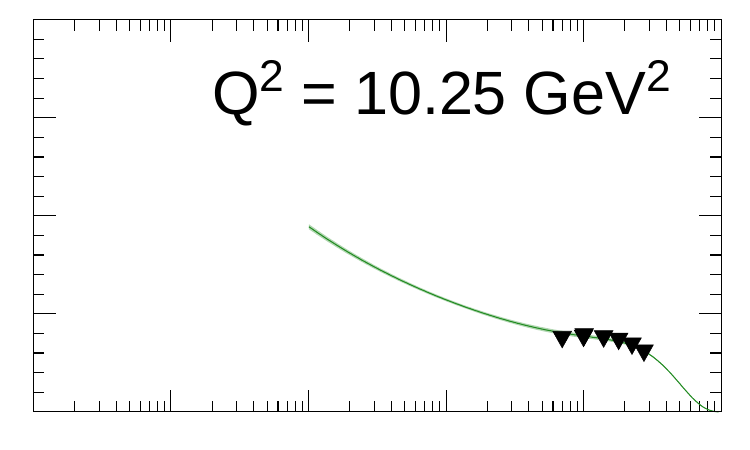}
\hspace*{-0.45cm} \includegraphics[width=2.8cm]{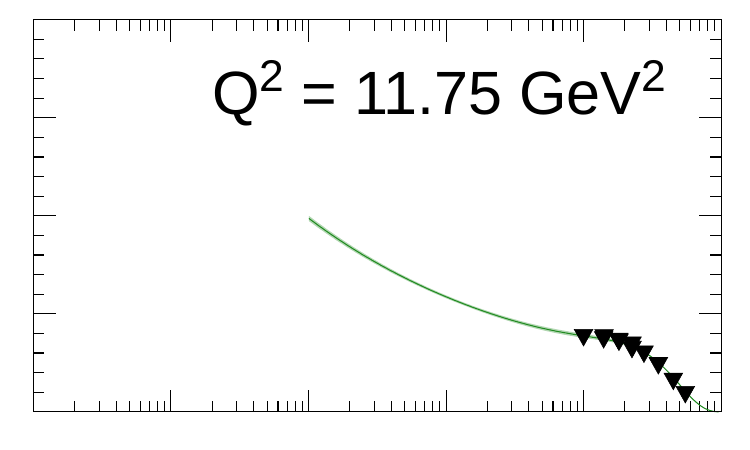}
\hspace*{-0.45cm} \includegraphics[width=2.8cm]{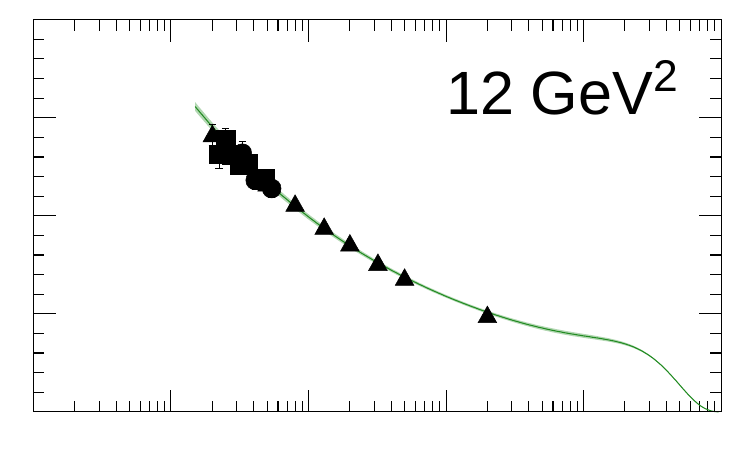}
\hspace*{-0.45cm} \includegraphics[width=2.8cm]{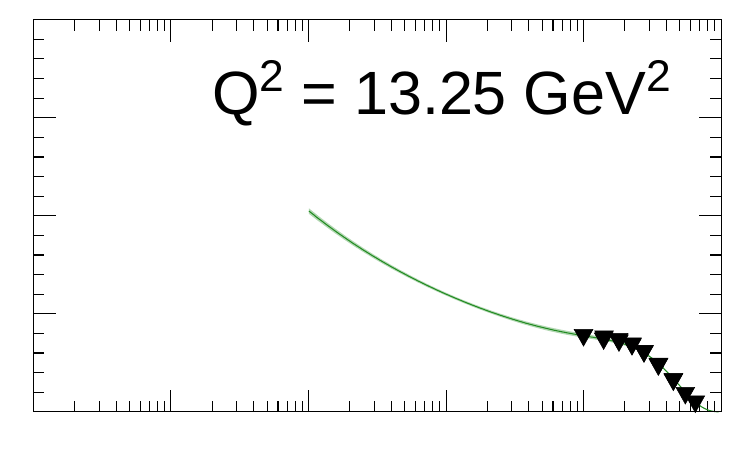}
\includegraphics[width=2.8cm]{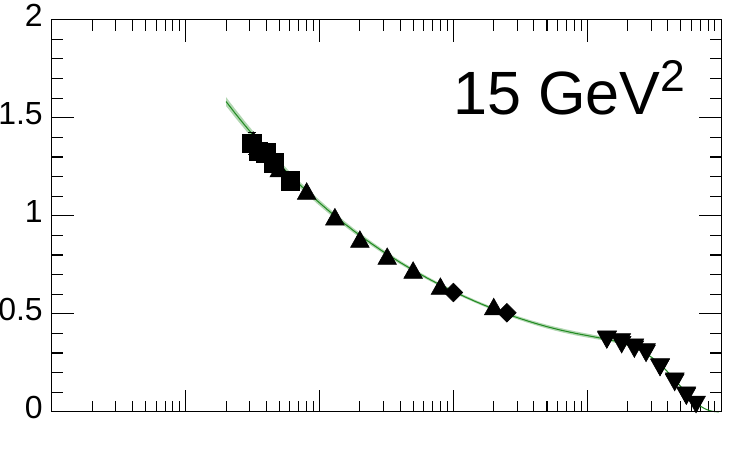}
\hspace*{-0.45cm} \includegraphics[width=2.8cm]{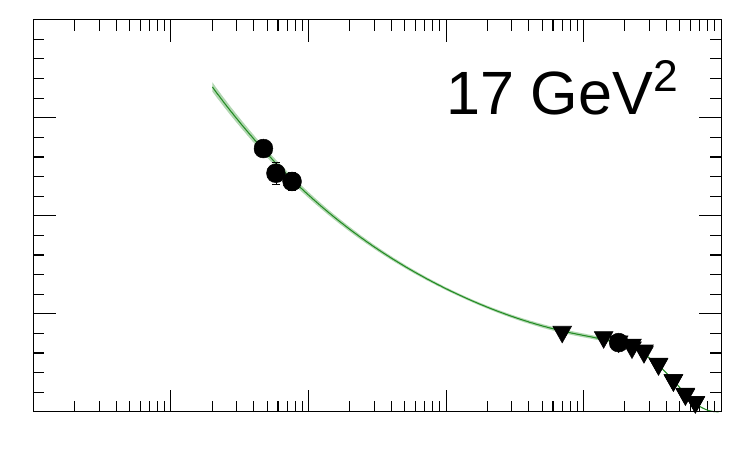}
\hspace*{-0.45cm} \includegraphics[width=2.8cm]{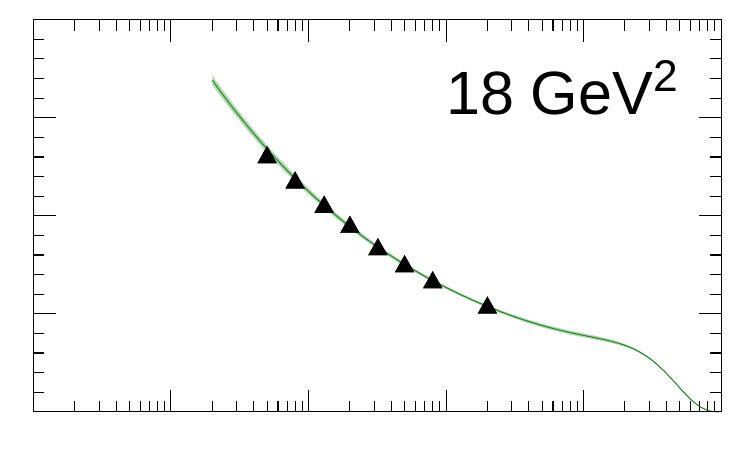}
\hspace*{-0.45cm} \includegraphics[width=2.8cm]{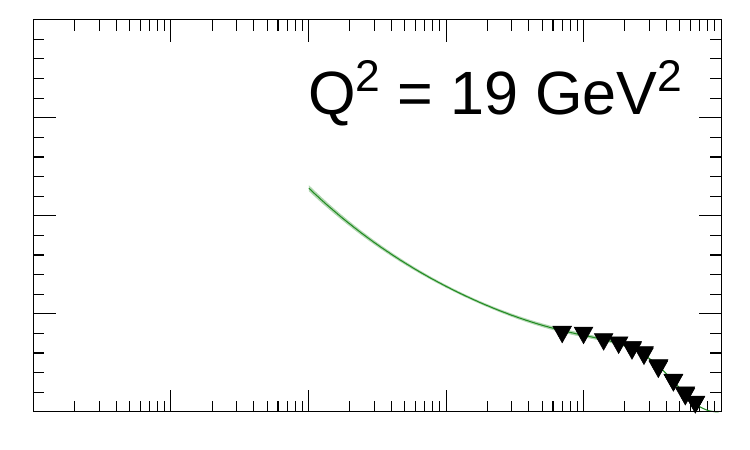}
\hspace*{-0.45cm} \includegraphics[width=2.8cm]{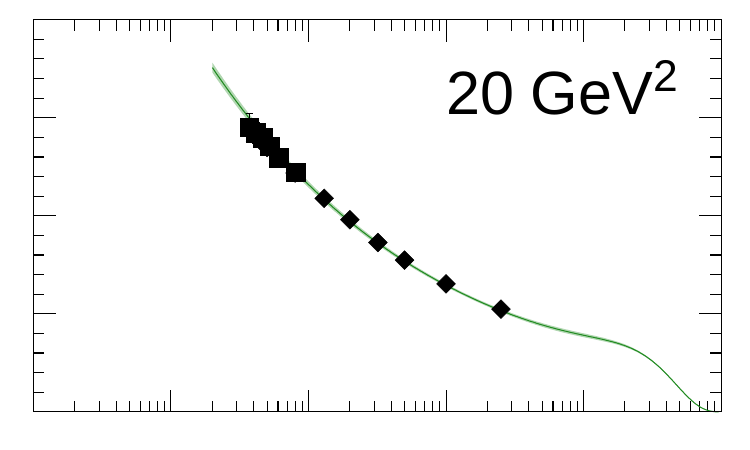}
\hspace*{-0.45cm} \includegraphics[width=2.8cm]{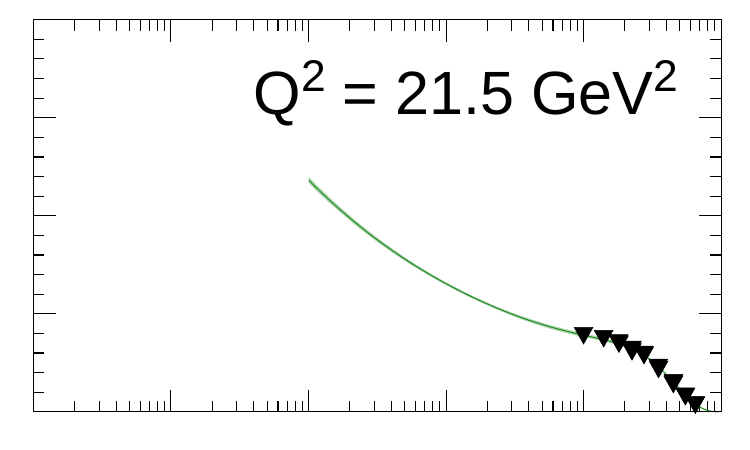}
\includegraphics[width=2.8cm]{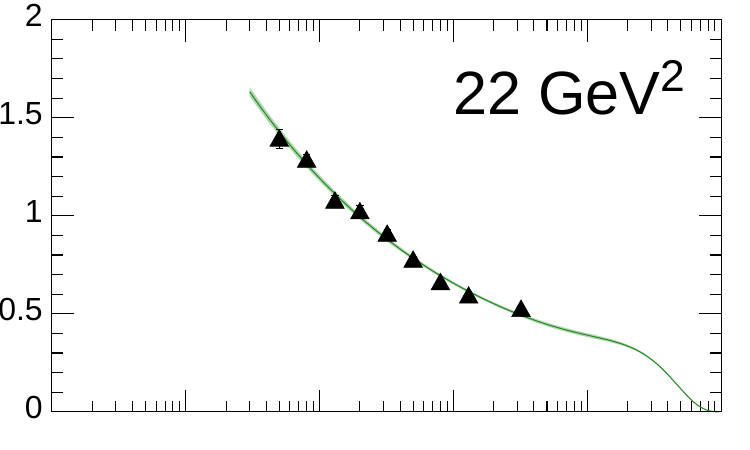}
\hspace*{-0.45cm} \includegraphics[width=2.8cm]{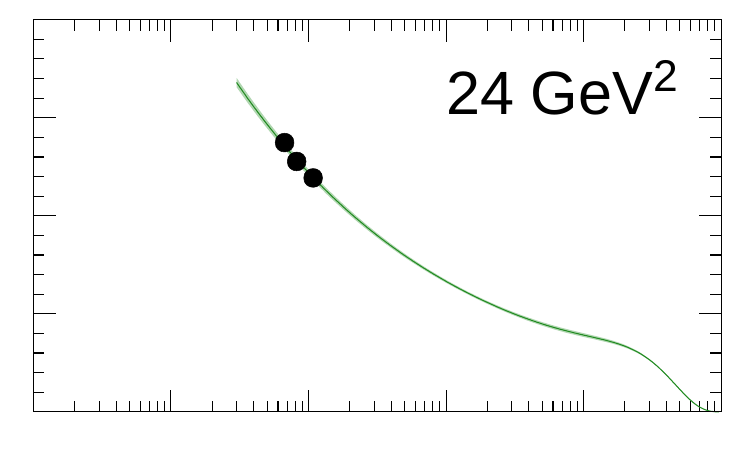}
\hspace*{-0.45cm} \includegraphics[width=2.8cm]{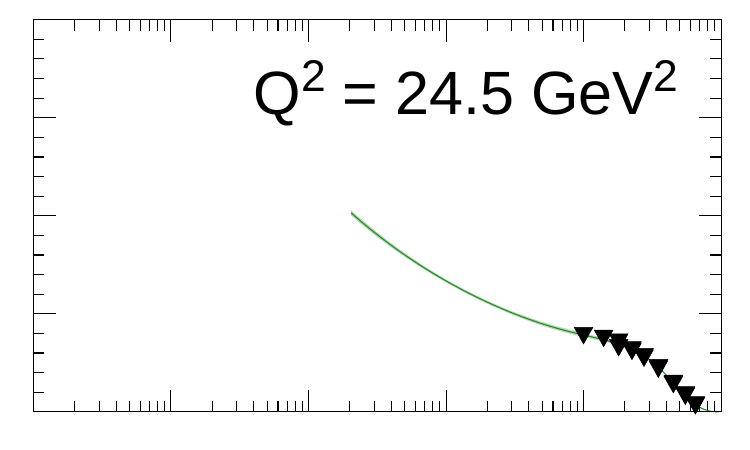}
\hspace*{-0.45cm} \includegraphics[width=2.8cm]{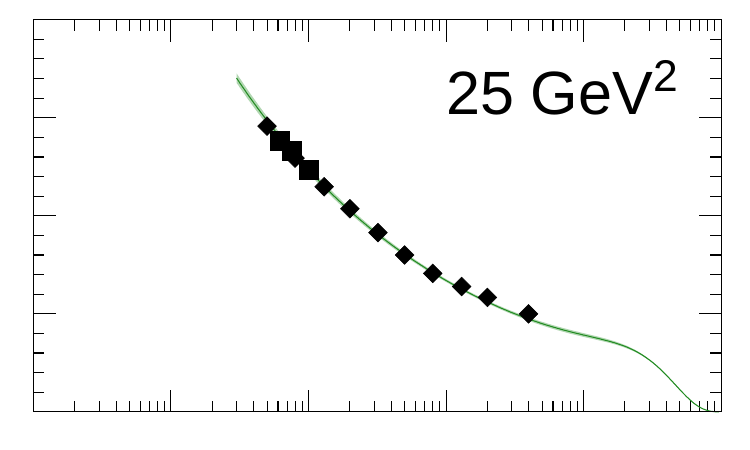}
\hspace*{-0.45cm} \includegraphics[width=2.8cm]{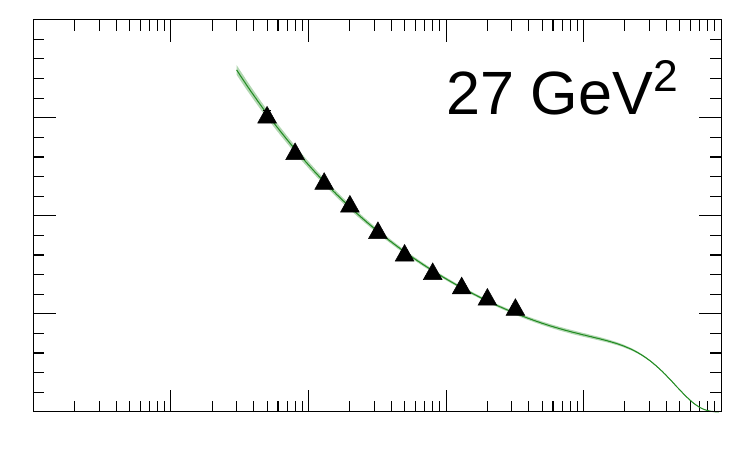}
\hspace*{-0.45cm} \includegraphics[width=2.8cm]{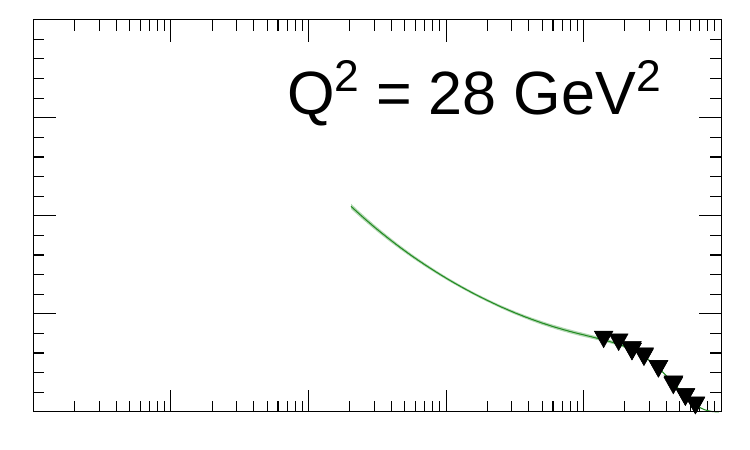}
\includegraphics[width=2.8cm]{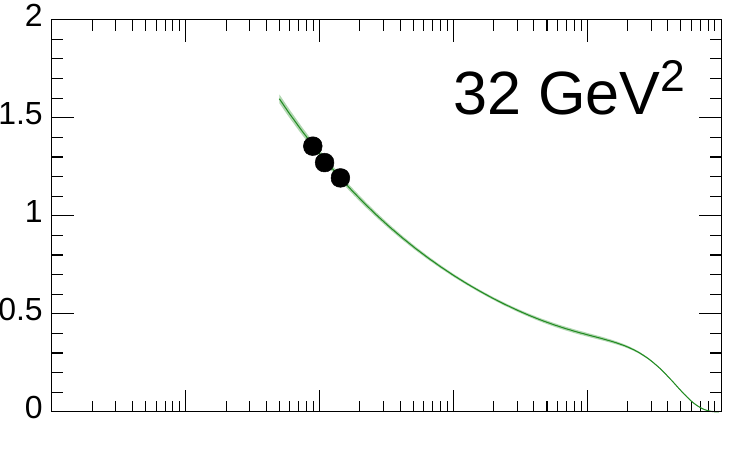}
\hspace*{-0.45cm} \includegraphics[width=2.8cm]{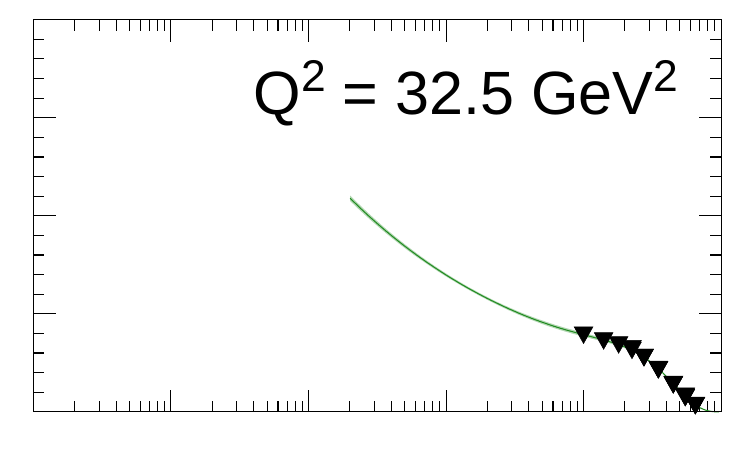}
\hspace*{-0.45cm} \includegraphics[width=2.8cm]{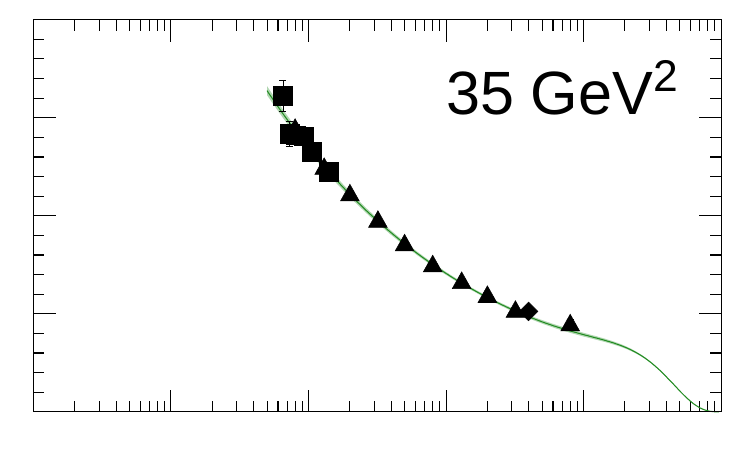}
\hspace*{-0.45cm} \includegraphics[width=2.8cm]{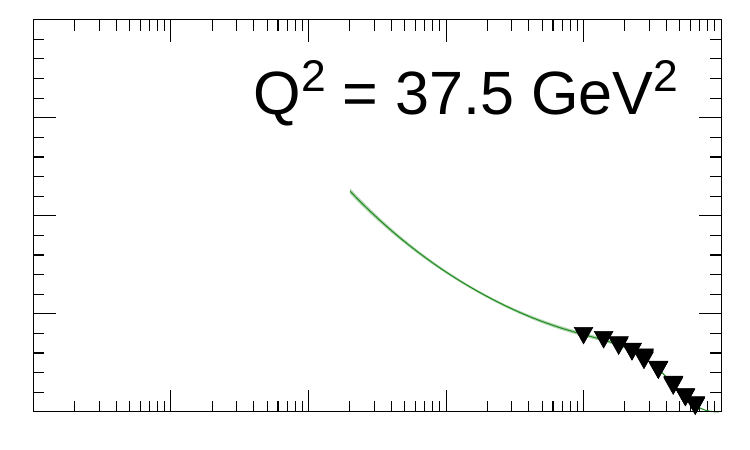}
\hspace*{-0.45cm} \includegraphics[width=2.8cm]{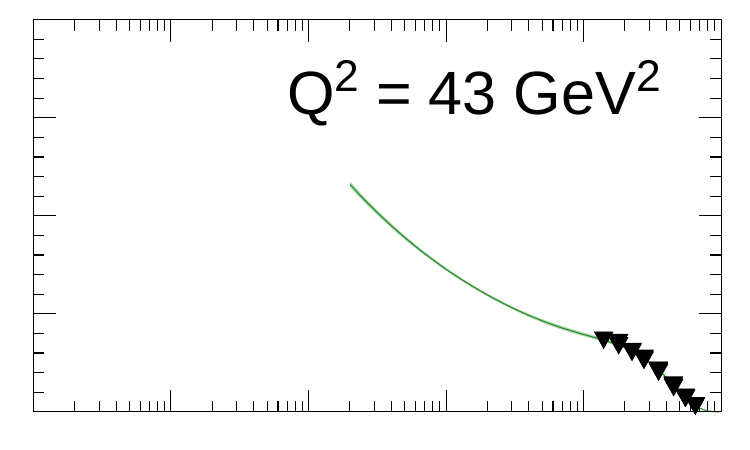}
\hspace*{-0.45cm} \includegraphics[width=2.8cm]{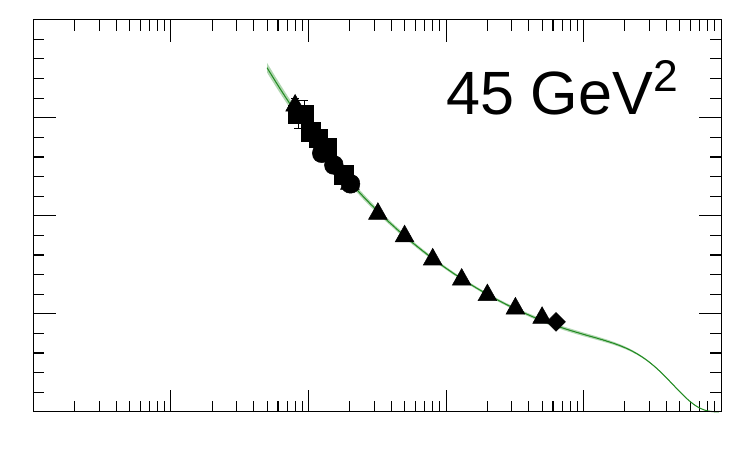}
\includegraphics[width=2.8cm]{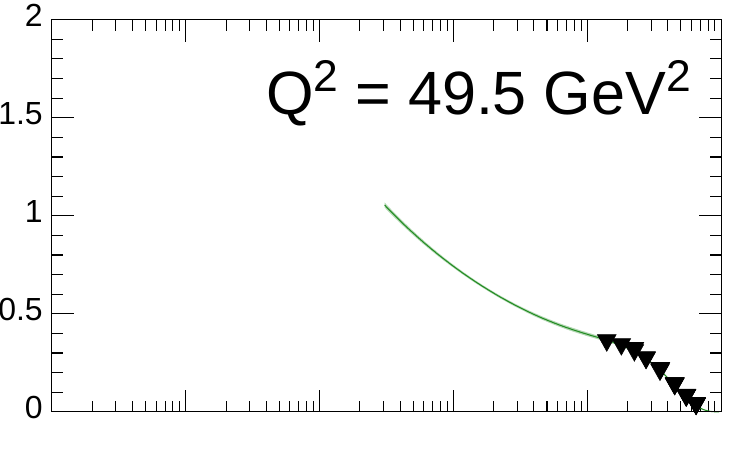}
\hspace*{-0.45cm} \includegraphics[width=2.8cm]{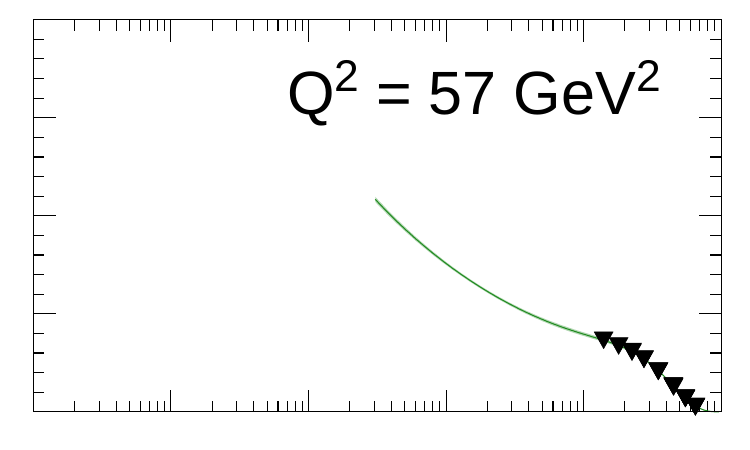}
\hspace*{-0.45cm} \includegraphics[width=2.8cm]{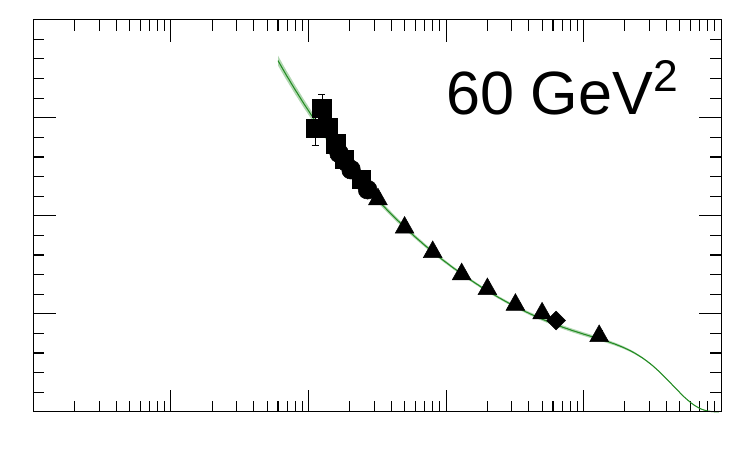}
\hspace*{-0.45cm} \includegraphics[width=2.8cm]{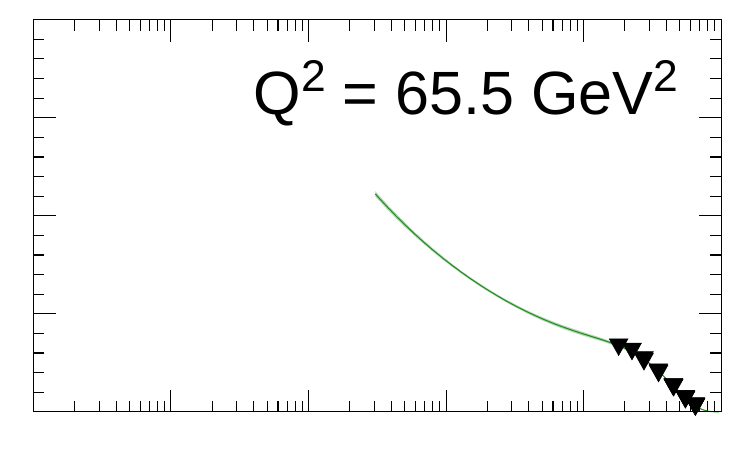}
\hspace*{-0.45cm} \includegraphics[width=2.8cm]{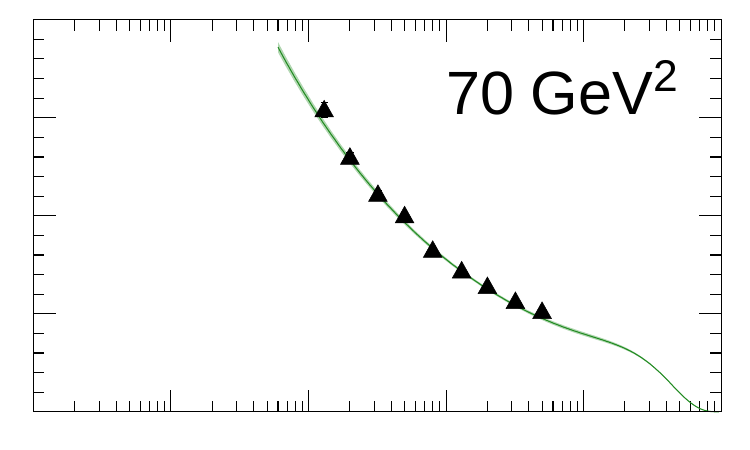}
\hspace*{-0.45cm} \includegraphics[width=2.8cm]{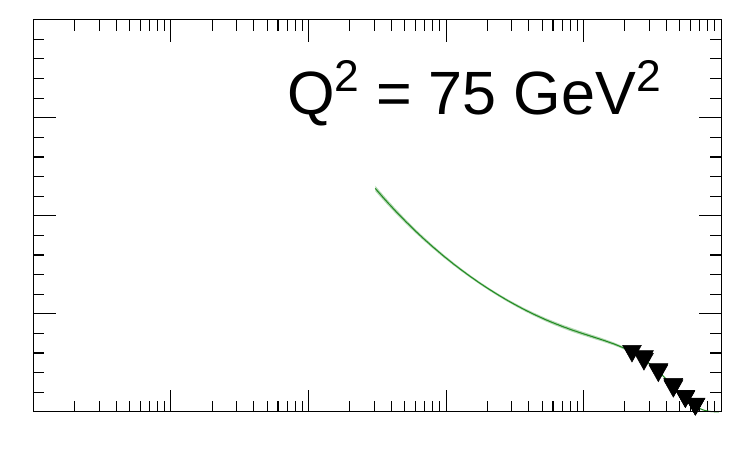}
\includegraphics[width=2.8cm]{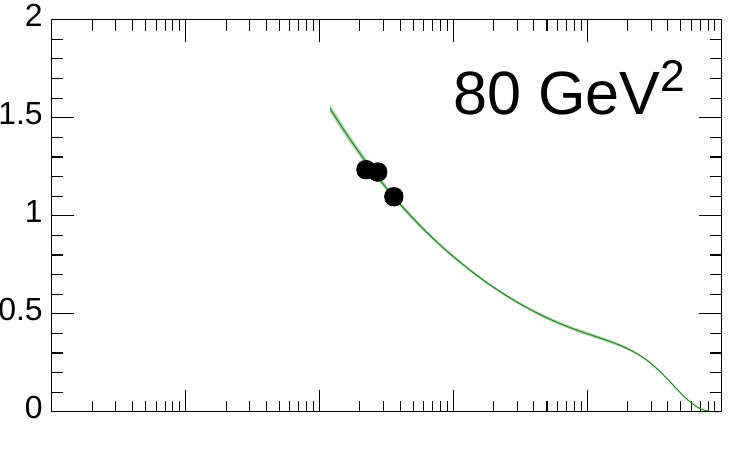}
\hspace*{-0.45cm} \includegraphics[width=2.8cm]{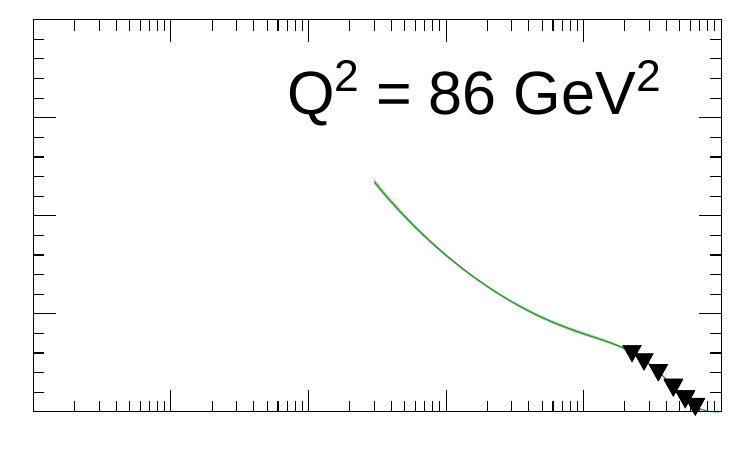}
\hspace*{-0.45cm} \includegraphics[width=2.8cm]{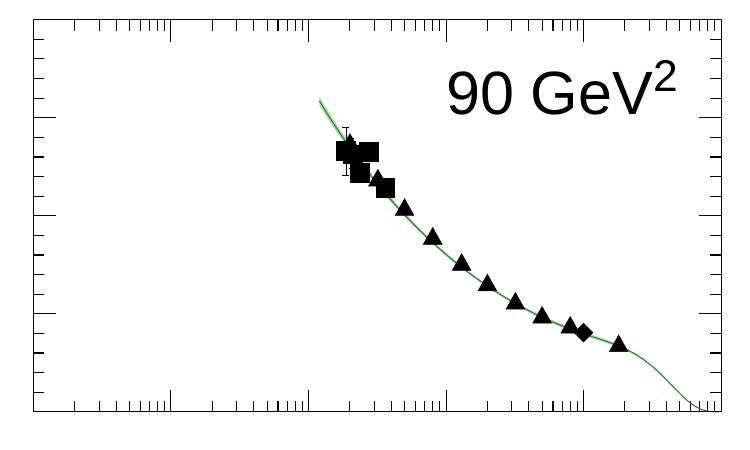}
\hspace*{-0.45cm} \includegraphics[width=2.8cm]{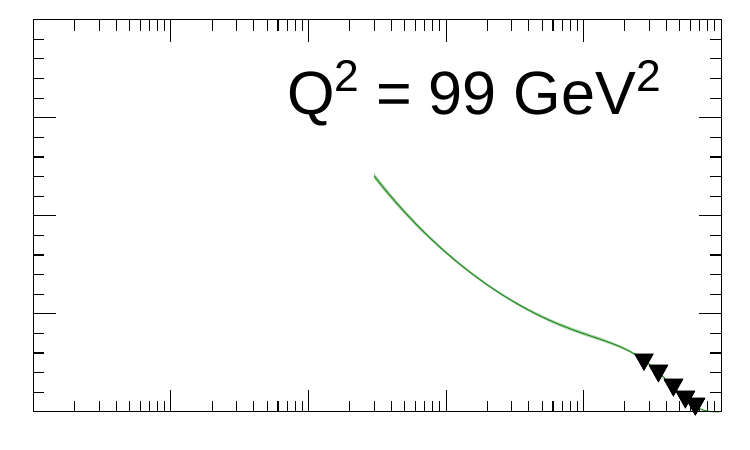}
\hspace*{-0.45cm} \includegraphics[width=2.8cm]{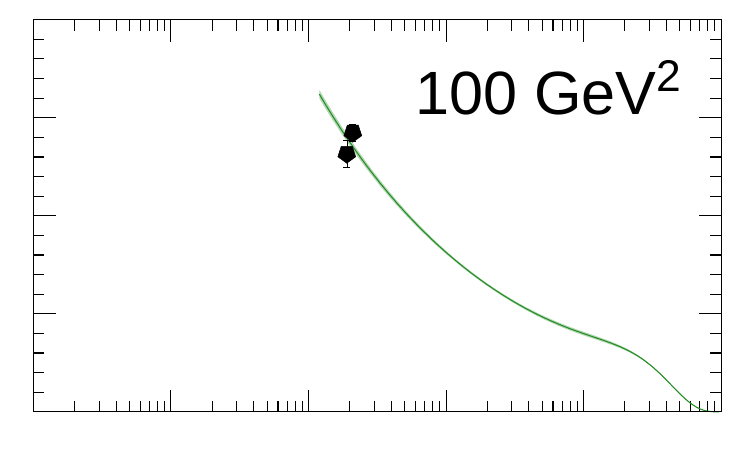}
\hspace*{-0.45cm} \includegraphics[width=2.8cm]{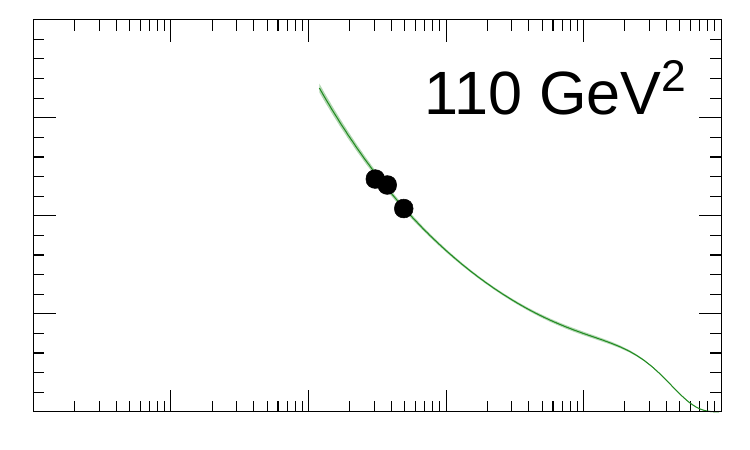}
\includegraphics[width=2.8cm]{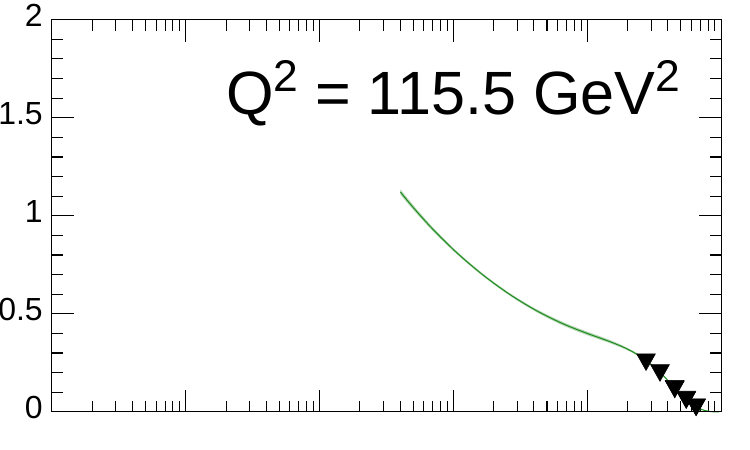}
\hspace*{-0.45cm} \includegraphics[width=2.8cm]{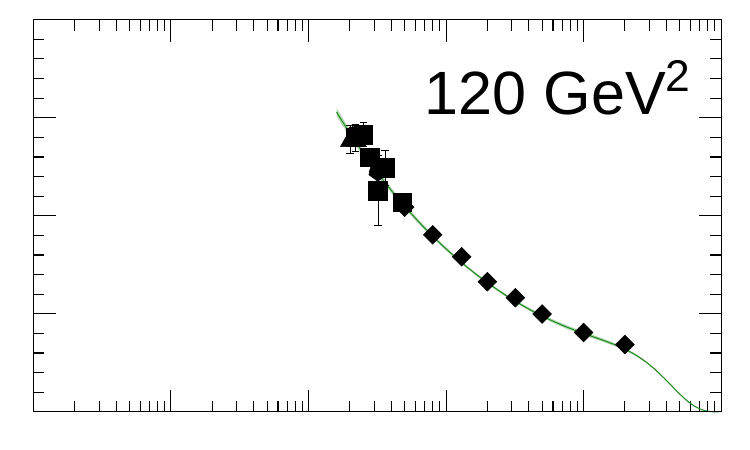}
\hspace*{-0.45cm} \includegraphics[width=2.8cm]{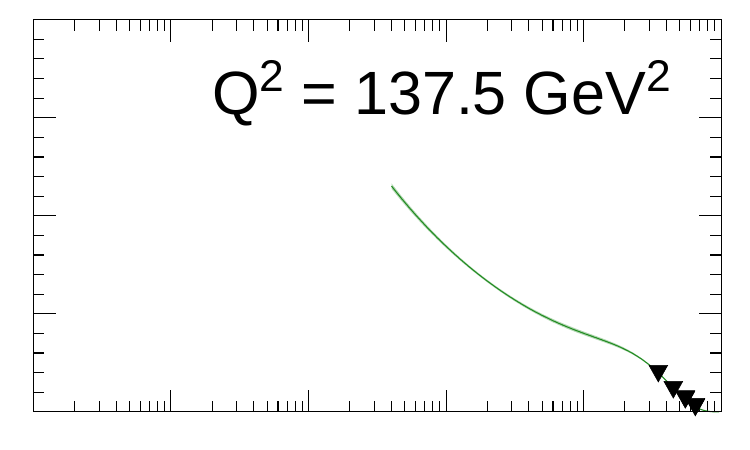}
\hspace*{-0.45cm} \includegraphics[width=2.8cm]{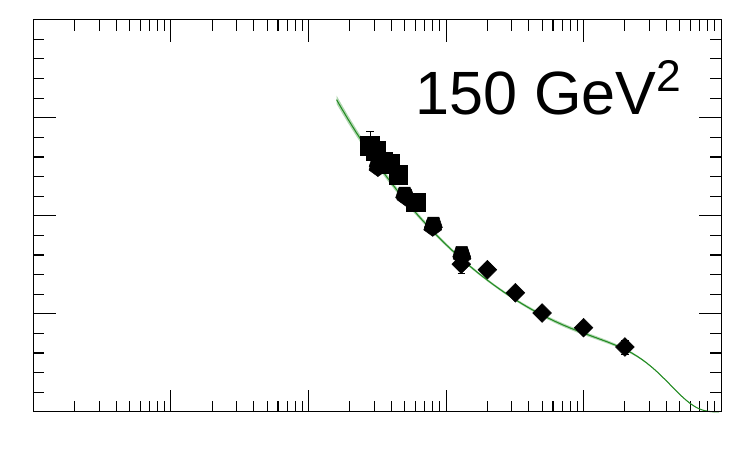}
\hspace*{-0.45cm} \includegraphics[width=2.8cm]{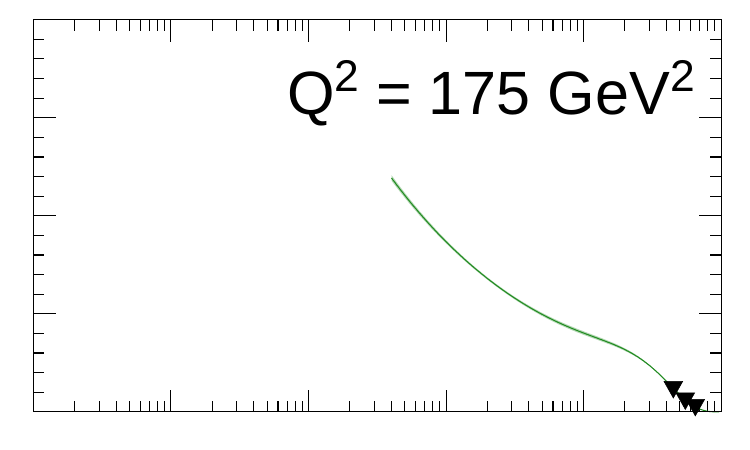}
\hspace*{-0.45cm} \includegraphics[width=2.8cm]{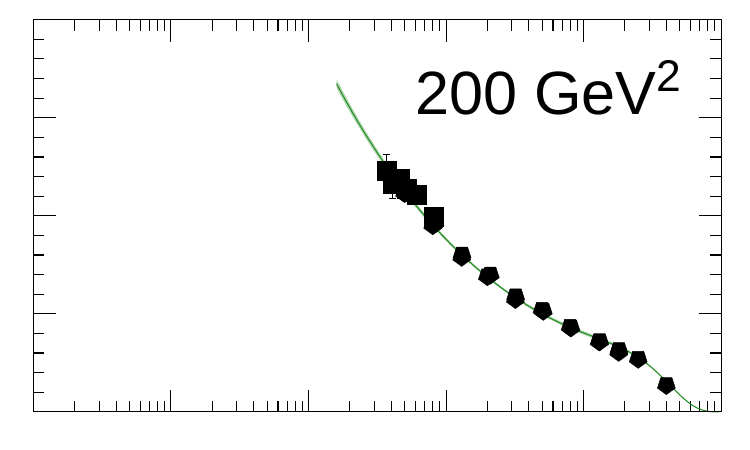}
\includegraphics[width=2.8cm]{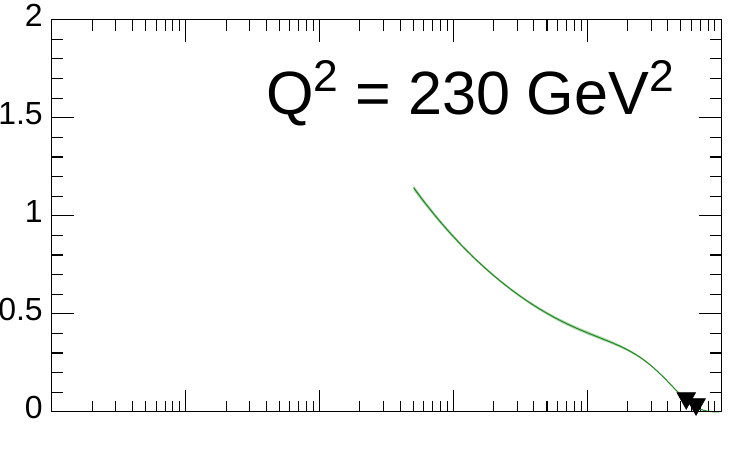}
\hspace*{-0.45cm} \includegraphics[width=2.8cm]{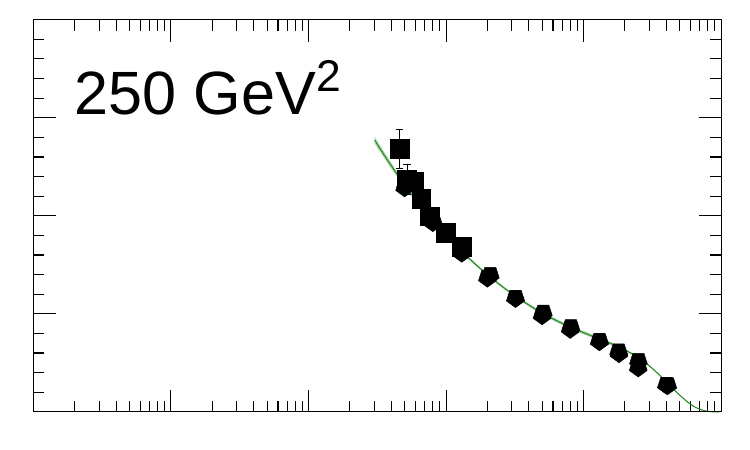}
\hspace*{-0.45cm} \includegraphics[width=2.8cm]{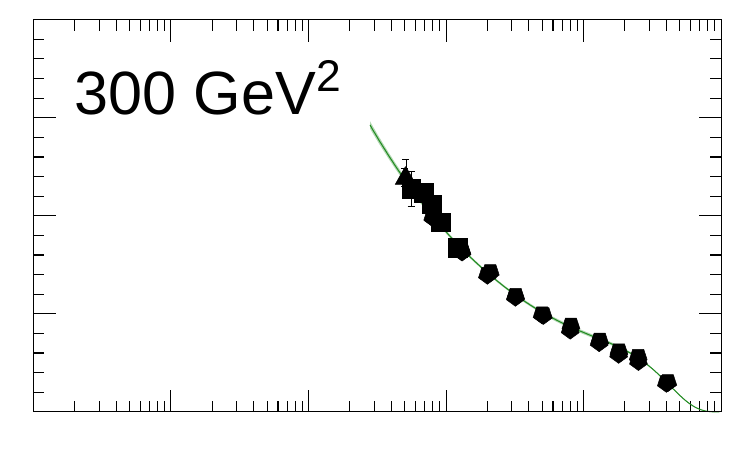}
\hspace*{-0.45cm} \includegraphics[width=2.8cm]{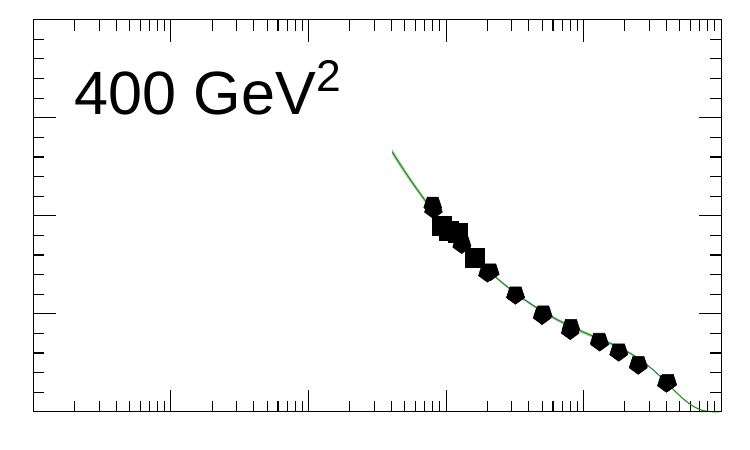}
\hspace*{-0.45cm} \includegraphics[width=2.8cm]{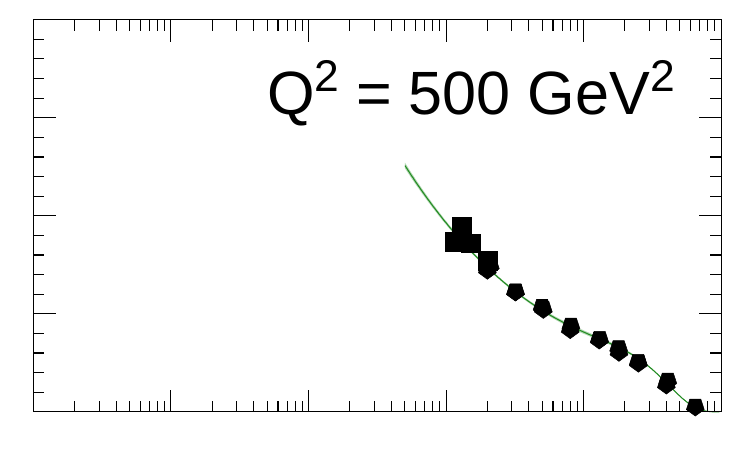}
\hspace*{-0.45cm} \includegraphics[width=2.8cm]{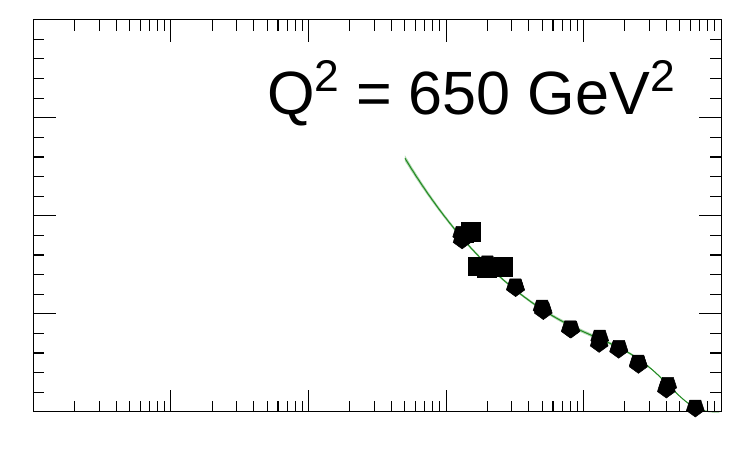}
\includegraphics[width=2.8cm]{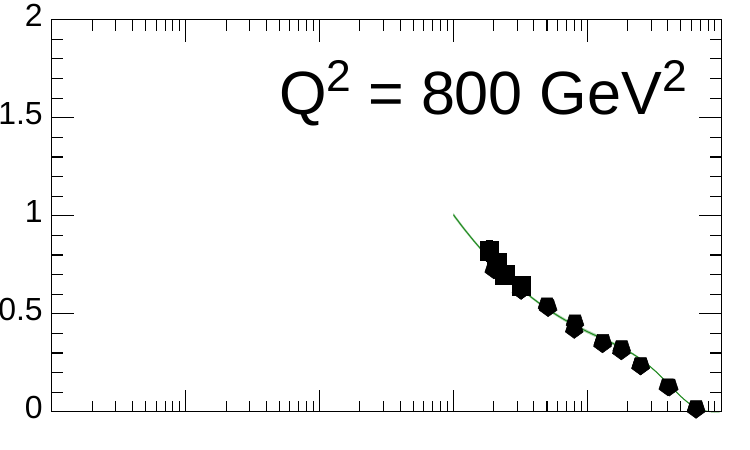}
\hspace*{-0.45cm} \includegraphics[width=2.8cm]{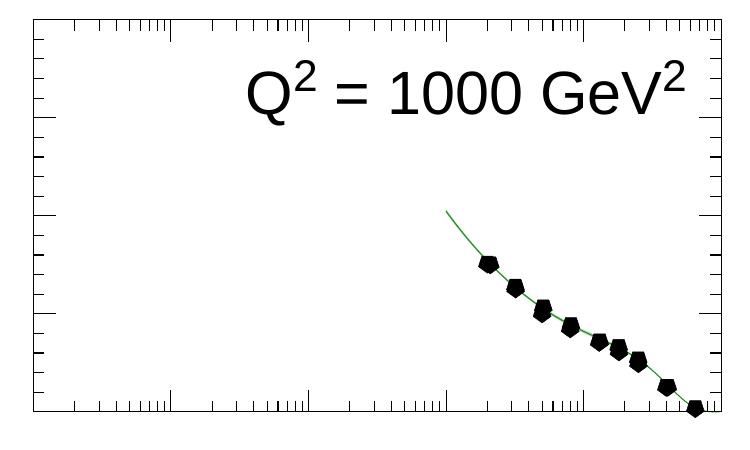}
\hspace*{-0.45cm} \includegraphics[width=2.8cm]{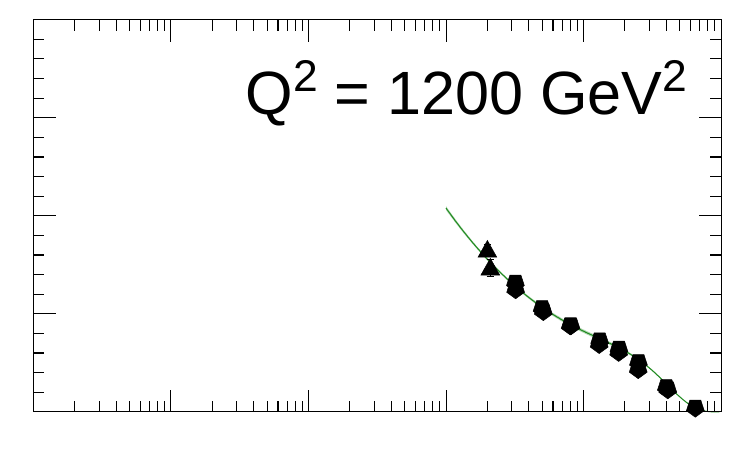}
\hspace*{-0.45cm} \includegraphics[width=2.8cm]{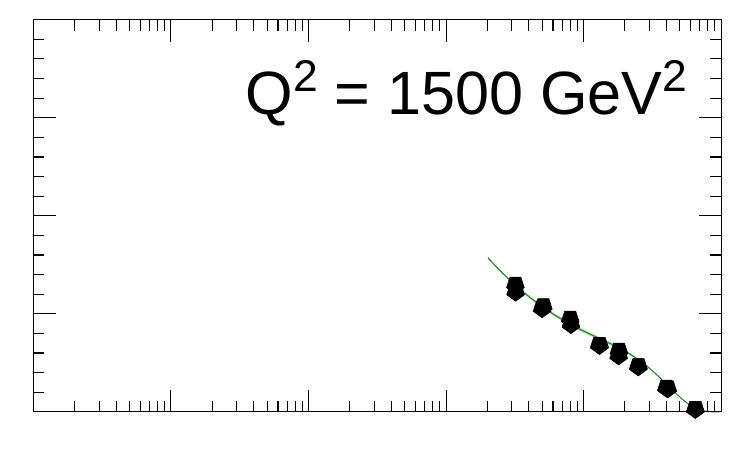}
\hspace*{-0.45cm} \includegraphics[width=2.8cm]{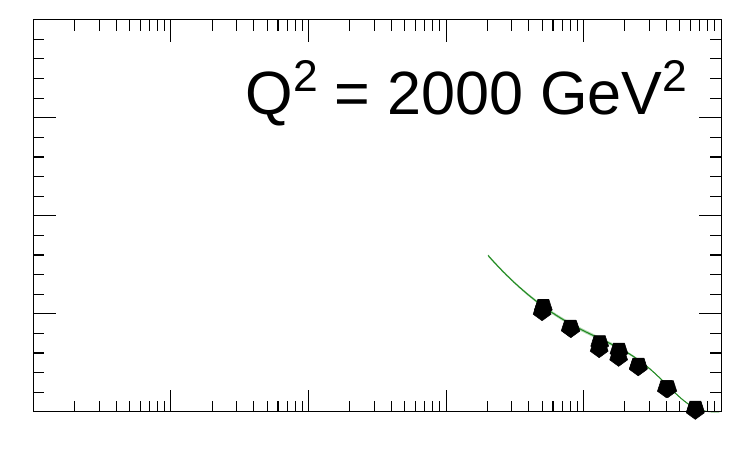}
\hspace*{-0.45cm} \includegraphics[width=2.8cm]{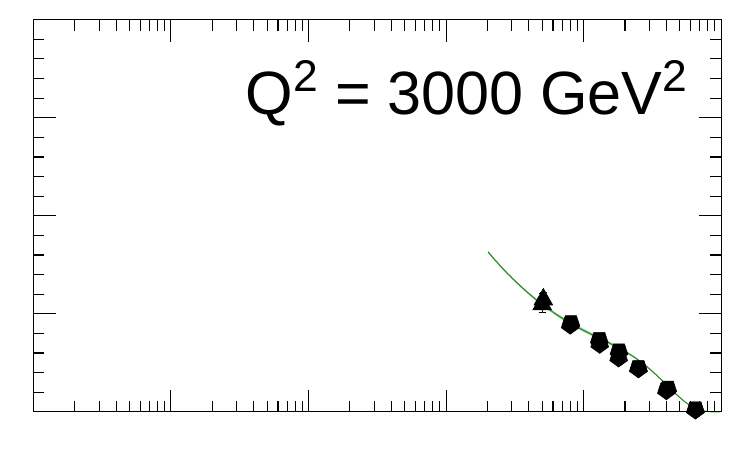}
\includegraphics[width=2.8cm]{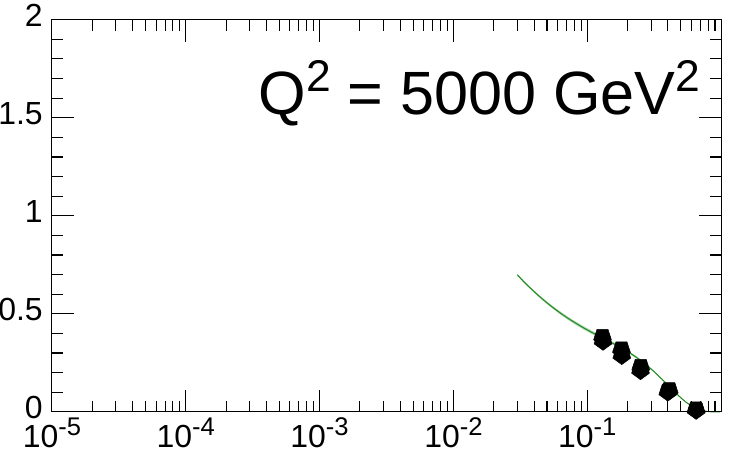}
\hspace*{-0.45cm} \includegraphics[width=2.8cm]{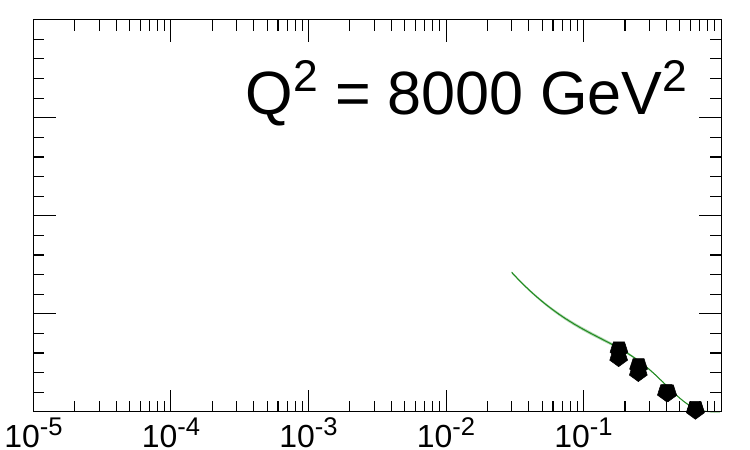}
\hspace*{-0.45cm} \includegraphics[width=2.8cm]{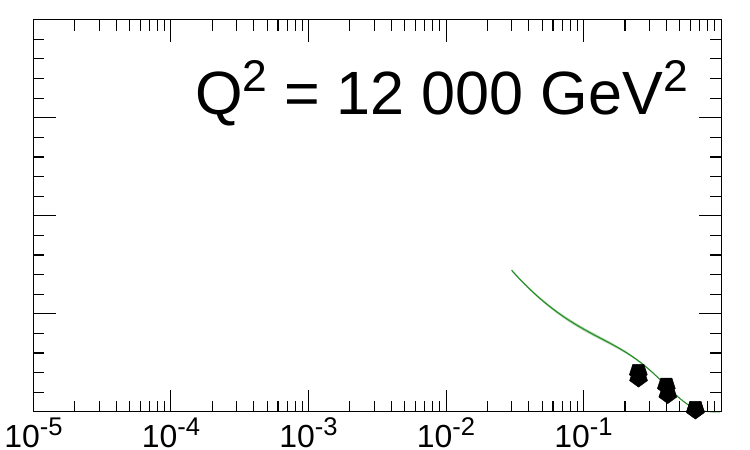}
\hspace*{-0.45cm} \includegraphics[width=2.8cm]{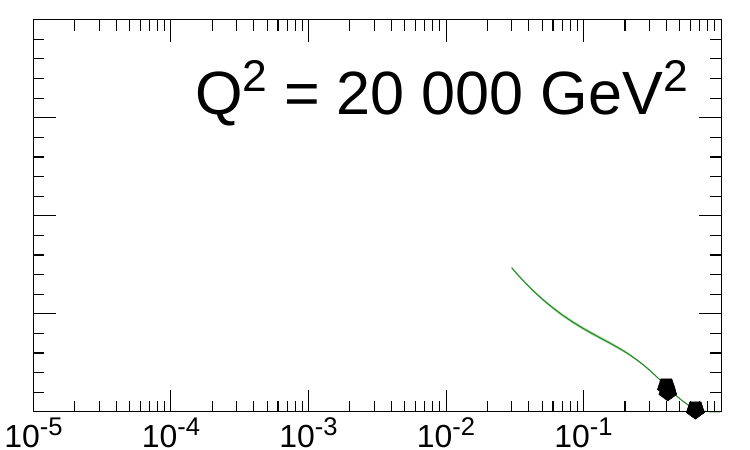}
\hspace*{-0.45cm} \includegraphics[width=2.8cm]{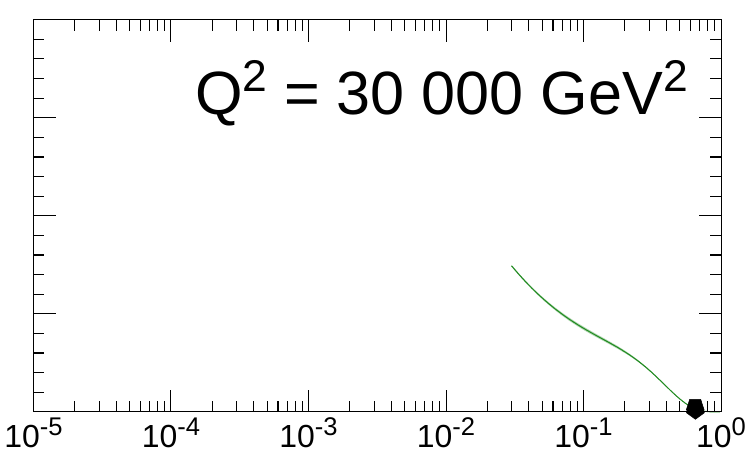}
\caption{Proton structure function $F_2(x, Q^2)$ as a function of $x$ for different $Q^2$.
The experimental data from BCDMS\cite{BCDMS-F2}, H1\cite{H1-F2-1,H1-F2-2,H1-F2-3,H1+ZEUS-F2}
and ZEUS\cite{H1+ZEUS-F2,ZEUS-F2} Collaborations are compared with our fits.}
\label{fig1}
\end{center}
\end{figure}

\begin{figure}
\begin{center}
\includegraphics[width=7.9cm]{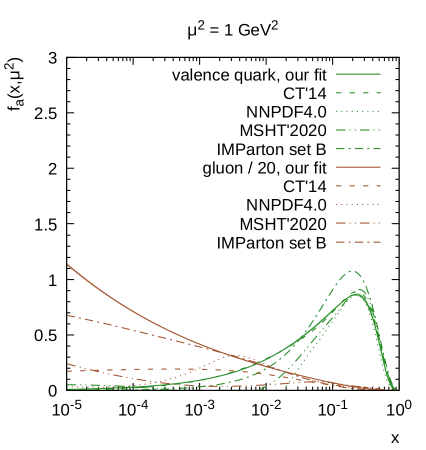}
\includegraphics[width=7.9cm]{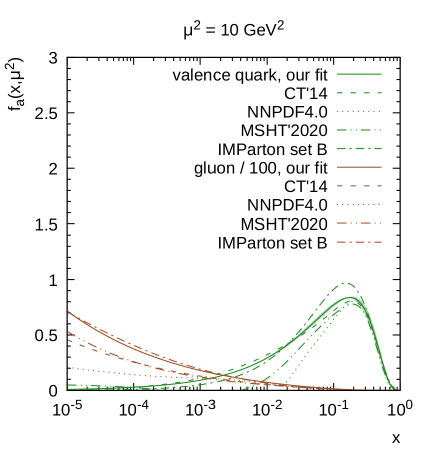}
\includegraphics[width=7.9cm]{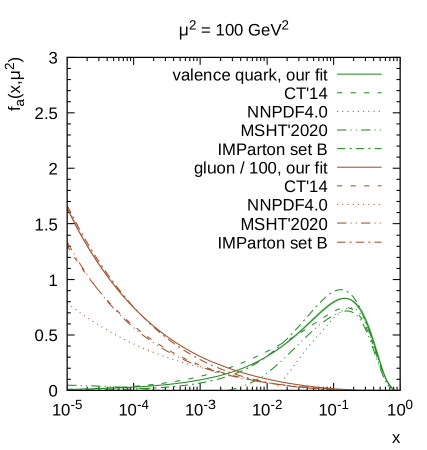}
\includegraphics[width=7.9cm]{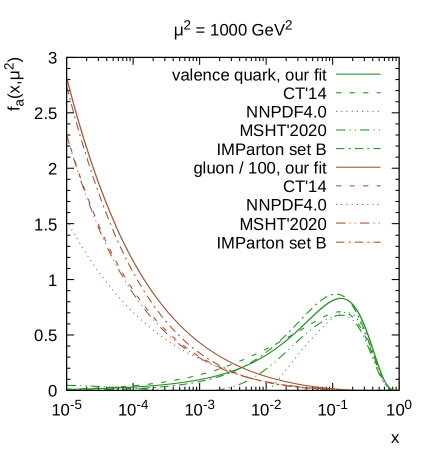}
\caption{Proton PDFs calculated as function of $x$ for different values of $\mu^2$.
For comparison we show here the results of numerical solutions of the DGLAP equations
performed by the CTEQ-TEA\cite{CT14}, NNPDF4.0\cite{NNPDF4}, MSHT'2020\cite{MSHT20} and IMP\cite{IMP} groups.}
\label{fig2}
\end{center}
\end{figure}

The results of our fit, comprising a total of $933$ points from $5$ data sets\cite{BCDMS-F2,H1-F2-1,H1-F2-2,H1-F2-3,H1+ZEUS-F2,ZEUS-F2},
are collected in Table~\ref{tbl:parameters4}.
A perfect goodness, $\chi^2/d.o.f. = 1.408$, is
achieved\footnote{We have obtained also $h_0 = - 0.65 \pm 0.14$, $h_1 = 0.44 \pm 0.13$ and $\bar h = 0.035 \pm 0.008$.}.
%with $N_f = 4$. %A somewhat worser, but still reasonable $\chi^2/d.o.f. = 3.616$ has been obtained for $N_f = 3$.
The experimental data involved into the fit procedures are
compared with our predictions in Fig.~\ref{fig1},
where shaded bands represent the fits uncertainties
summed in quadratures.
One can see that good agreement is obtained in a wide range of $x$ and $Q^2$.
%for both $N_f = 3$ and $N_f = 4$.

We note that newly fitted values of $A_q$ and $A_g$
%slightly
differ from the previous results\cite{PDFs-our-previous-2}. The main sourse of this difference is
that the small-$x$ PDF asymptotics have been used in the analysis\cite{PDFs-our-previous-2} and
experimental data on proton structure function $F_2(x,Q^2)$ were considered at low $x$ and not very high $Q^2$ only.
In contrast, here we extended the consideration into whole region of $x$ and $Q^2$
and take into account all the available data sets.
Thus, in this point our analysis significantly improves the previous consideration\cite{PDFs-our, PDFs-our-previous-2}.

\begin{figure}
\begin{center}
\includegraphics[width=7.9cm]{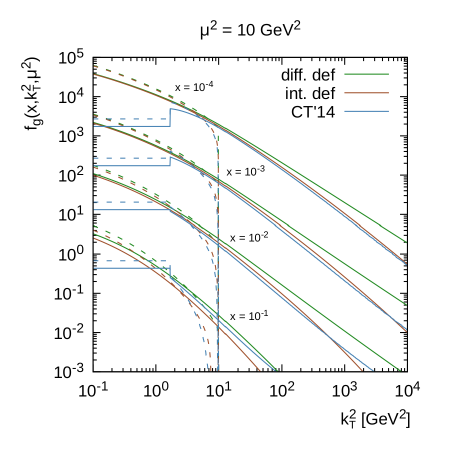}
\includegraphics[width=7.9cm]{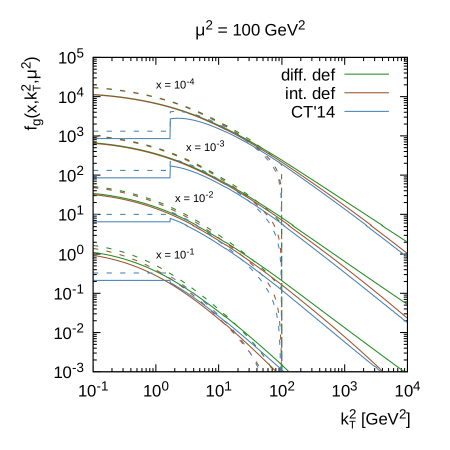}
\includegraphics[width=7.9cm]{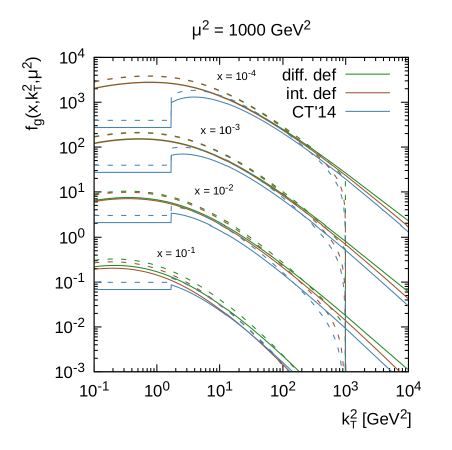}
\includegraphics[width=7.9cm]{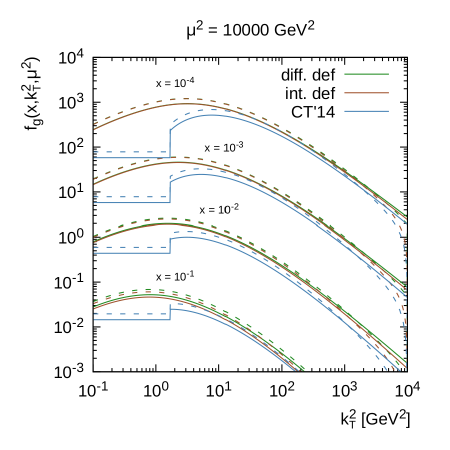}
\caption{TMD gluon densities in a proton calculated as functions of ${\mathbf{k}_T^2}$ for different $x$ and $\mu^2$
using the angular ordering (solid curves) and strong ordering (dashed curves) conditions.
Both the differential and integral KMR/WMR definitions have been applied for $N_f = 4$.
For comparison we show corresponding results obtained numerically employing~(\ref{eq-KMR-int}), where
conventional PDFs from CT'14 set\cite{CT14} have been used as an input.}
\label{fig3}
\end{center}
\end{figure}

Some of the derived PDFs are shown in Fig.~\ref{fig2} as functions of $x$
for several scales $\mu^2$, namely, $\mu^2 = 1$, $10$, $100$ and $1000$ GeV$^2$.
%The green and brown curves
%correspond to fits performed with $N_f = 3$ and $N_f = 4$, respectively.
Additionally, we show here the comparison between our predictions and
corresponding FFNS results obtained by the
CTEQ-TEA Collaboration\cite{CT14} at $N_f = 4$ and VFNS
%variable-flavor-number-sheme (VFNS)
predictions
of the NNPDF4.0\cite{NNPDF4}, MSHT'2020\cite{MSHT20}
and IMP\cite{IMP} groups.
One can see that, in fact, up to now the sizeble discrepancy
between the different PDFs could be seen,
especially at low $\mu^2 \sim 1$~GeV$^2$.
Nevertheless, at moderate and higher $\mu^2$
we find a reasonable agreement between our analytical results
and relevant numerical analyses.
In particular, our fit and VFNS
predictions from IMP group are very close to each other.

In Fig.~\ref{fig3} we show the TMD gluon densities in a proton
calculated according to~(\ref{uPDF}) and (\ref{eq1-Sg_kt}) --- (\ref{uPDF2.1g}) with different treatment of the
angular ordering condition~(\ref{eq-deltas})
%(angular and strong ordering)
as functions of transverse momentum $\mathbf{k}_T^2$
for several values of $x$ and scale $\mu^2$.
%These results are based on the fit performed with $N_f = 4$.
The solid and dashed curves correspond to the results obtained with the
angular and strong ordering conditions, respectively.
One can see that the differential and integral definitions of the
KMR/WMR procedure lead to very close predictions
at small $x$ and low and moderate transverse momenta $\mathbf{k}_T^2$,
whereas some differences
between them occur at large $x$ and/or large $\mathbf{k}_T^2$.
It fully coincides with the observations\cite{KMR-discussion-1},
where the necessity of using the cutoff $\Delta$ dependent PDFs %as an input
in the case of differential KMR/WMR definition was pointed out (see also discussions\cite{KMR-discussion-3, KMR-discussion-4}).
Additionally, we plot here the TMD gluon densities obtained
numerically employing the general integral KMR/WMR definition formula~(\ref{eq-KMR-int}), where
the conventional PDFs from CT'14 set\cite{CT14} have been used as an input
and popular normalization condition~(\ref{eq-norm}) with $\mu^2_{\rm max} = \mu^2$
and $Q_0 = 1.3$~GeV was applied.
We find a notable agreement between
these results and our corresponding predictions derived
according to analytical expressions~(\ref{eq1-Sg_kt}) --- (\ref{uPDF2.1g})
at $\mathbf{k}_T^2 > Q_0^2$.
Note that the flat behavior of former
at $\mathbf{k}_T^2 < Q_0^2$ is due to commonly used assumptions
%for TMDs
in low $\mathbf{k}_T^2$ region.
Finally, we demonstrate here again
that the strong ordering condition leads to a steep drop of the gluon
densities beyond the hard scale $\mu^2$.
It contrasts with the angular ordering, where the gluon transverse momentum
is allowed to be larger than $\mu^2$ (see \cite{KMR-LO}).

\subsection{Testing the TMDs: beauty production in $pp$ collisions} \indent

Now we test the obtained analytical expressions~(\ref{uPDF}) and (\ref{eq1-Sg_kt}) --- (\ref{uPDF2.1g})
with beauty production in $pp$ collisions at the LHC.
These processes are promising for such study since
they are known to be strongly sensitive to the gluon content of
the proton\footnote{The derived TMD quark densities in a proton
can be tested, for example, with Dell-Yan processes, where quark distributions
play a leading role. However, we leave it for the dedicated study.}.

Our calculations are performed in the FFNS scheme
with taking into account TMD gluon dynamics in a proton
and strongly based on the previous considerations\cite{bb-kt-our1, bb-kt-our2}.
In this approach, the main contribution to the heavy flavor
production cross section comes from the off-shell gluon-gluon fusion
subprocess $g^*(k_1) + g^*(k_2) \to Q(p_1) + \bar Q(p_2)$,
where $Q$ denotes outgoing beauty quark and
four-momenta of all the particles are indicated in parentheses.
The contribution from the quark-antiquark annihilation is of almost no importance
because of comparatively low quark densities.
According to the high energy factorization prescription\cite{HighEnergyFactorization, kt-factorization}, corresponding production cross
section can be written as a convolution
of the TMD gluon densities and off-shell (depending on the transverse
momenta of the incoming off-mass shell gluons) hard scattering amplitudes.
The latter has been calculated a long time ago (see, for example,\cite{HighEnergyFactorization, kt-factorization, bb-kt-our1}).
The detailed description of the calculation steps (including the evaluation of the off-shell
gluon-gluon amplitudes) can be found\cite{bb-kt-our1}.
We only specify here the essential numerical parameters. So,
following\cite{PDG}, we set the %charm and
beauty quark mass
%$m_c = 1.5$~GeV and
$m_b = 4.75$~GeV.
Then, as it is often done, we choose the default renormalization
and factorization scales $\mu_R$ and $\mu_F$ to
be equal to the leading heavy jet transverse momentum.
The calculations were performed using Monte-Carlo event generator \textsc{pegasus}\footnote{{\texttt https://theory.sinp.msu.ru/doku.php/pegasus/news}}\cite{PEGASUS}.

%First we discuss the inclusive $b$-jet production at the LHC.
The experimental data for beauty production in $pp$ collisions at the LHC come from both CMS and ATLAS
Collaborations.
So, the CMS Collaboration has measured the double differential cross section $d\sigma/dp_Tdy$
in the range $18 < p_T < 196$~GeV at $\sqrt s = 7$~TeV in five $b$-jet rapidity regions, namely, $|y(b)| < 0.5$,
$0.5 < |y(b)| < 1$, $1 < |y(b)| < 1.5$, $1.5 < |y(b)| < 2$ and $2 < |y(b)| < 2.2$ as a
function of the leading $b$-jet transverse momentum\cite{bjet-CMS-7}.
In the ATLAS analysis\cite{bjet-ATLAS-7}, the inclusive $b$-jet cross section was measured as a
function of transverse momentum $p_T$ in the range $20 < p_T(b) < 400$~GeV and four
rapidity subdivisions, $|y(b)| < 0.3$, $0.3 < |y(b)| < 0.8$,  $0.8 < |y(b)| < 1.2$ and
$1.2 < |y(b)| < 2.1$. In addition, the $b\bar b$-dijet cross section was measured as a
function of the dijet invariant mass $M$ in the range $110 < M < 760$~GeV, azimuthal angle
difference $\Delta \phi$ between the two leading $b$-jets and angular variable $\chi = \exp |y_1 - y_2|$
for jets with $p_T > 40$~GeV in two dijet mass regions, $110 < M < 370$~GeV and $370 < M < 850$~GeV.

\begin{figure}
\begin{center}
\includegraphics[width=7.7cm]{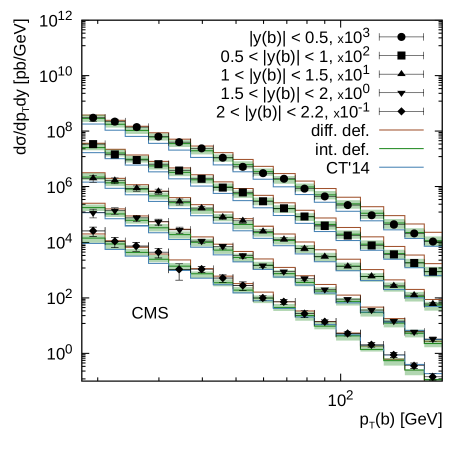} \hspace*{0.3cm}
\includegraphics[width=7.7cm]{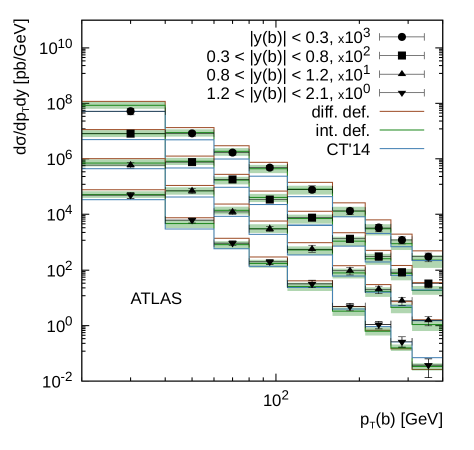}
\caption{The transverse momentum distributions of inclusive $b$-jet
production at $\sqrt s = 7$~TeV as functions of the leading jet transverse momentum in
different rapidity regions. The kinematical cuts are described in the text.
The experimental data are from CMS\cite{bjet-CMS-7} and ATLAS\cite{bjet-ATLAS-7}.}
\label{fig4}
\end{center}
\end{figure}

\begin{figure}
\begin{center}
\includegraphics[width=7.7cm]{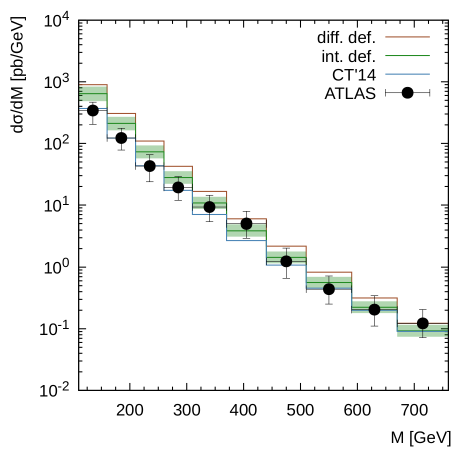} \hspace*{0.3cm}
\includegraphics[width=7.7cm]{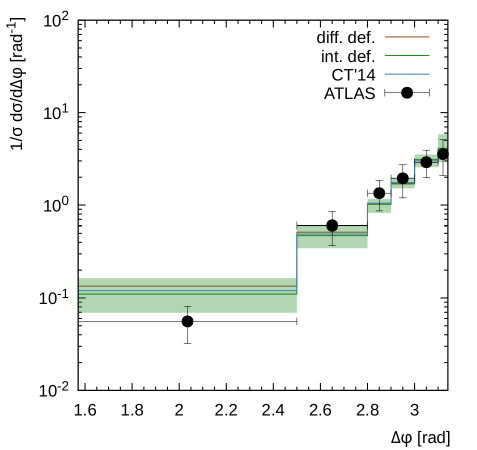}
\includegraphics[width=7.7cm]{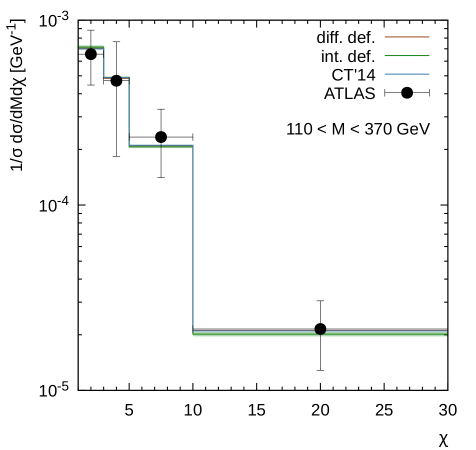} \hspace*{0.3cm}
\includegraphics[width=7.7cm]{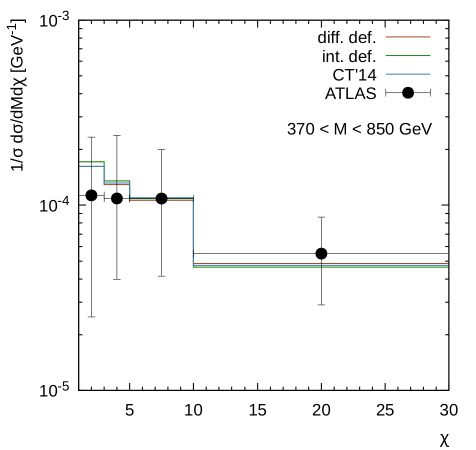}
\caption{The $b\bar b$-dijet invariant mass $M$, azimuthal angle difference $\Delta \phi$ and $\chi$
distributions of $b\bar b$ production at $\sqrt s = 7$~TeV. The kinematical cuts are described in the text.
The experimental data are from ATLAS\cite{bjet-ATLAS-7}.}
\label{fig5}
\end{center}
\end{figure}

The results of our calculations are shown in Figs.~\ref{fig4} and \ref{fig5} in comparison with
experimental data\cite{bjet-CMS-7, bjet-ATLAS-7}.
Here we plot predictions based on both differential and integral KMR/WMR definitions (\ref{uPDF}) and (\ref{eq1-Sg_kt}) --- (\ref{uPDF2.1g})
with the AO condition applied.
Additionally, we show the results obtained by employing the
integral formula~(\ref{eq-KMR-int}), where
the conventional PDFs from CT'14 set\cite{CT14} have been used as an input, as it was done earlier (see Fig.~\ref{fig3}).
Solid histograms corresponds to our central results, where we fixed
both renormalization and factorization scales at their default values.
The shaded bands correspond to scale uncertainties of our calculations
with (\ref{eq1-Sg_kt}) --- (\ref{uPDF2.1g}).
As usual, the latter were estimated by varying the scales $\mu_R$ and $\mu_F$ by a factor of $2$
around their default values.
We find that good agreement with the LHC data on the $b$-jet transverse
momenta is achieved with integral KMR/WMR definition in each of the rapidity subdivisions, both in normalization and shape (green histograms).
In contrast, the predictions based on the differential definition overestimate
the data at central rapidities and large transverse momenta,
thus leading to a worse description of the data (see Fig.~\ref{fig4}).
However, these predictions are close to the data in forward kinematical region in the whole $p_T$ range.
At low and moderate transverse momenta results obtained with both KMR/WMR
definitions practically coincide.
The predictions based on conventional PDFs from CT'14 set
tend to underestimate the data at low $p_T \sim 20$~GeV and
overestimate the data at large $p_T \sim 400$~GeV at forward rapidities.
The observed difference in predictions for $b$-jet transverse momentum
distributions is an immediate consequence of the difference between these
gluon densities at large $\mathbf{k}_T^2$ shown in Fig.~\ref{fig3}.
So, the general conclusion\cite{KMR-discussion-1}
that integral KMR/WMR definition with ordinary PDFs used as an input
is more preferable for phenomenological application is confirmed.
Specially we point out a
good description of the LHC data at large $p_T$,
where the essentially large $x$ region is probed.
That is due to accurate taking into account
large-$x$ asymptotics of solutions of the DGLAP equations
in our present calculations.
Note that in our previous consideration\cite{TMDs-KMR-our-1}
%\cite{KMR-DAS-1}
such extension of the TMD parton densities to the large-$x$
region has been modelled.

Both the considered above analytical expressions (\ref{uPDF}) and (\ref{eq1-Sg_kt}) --- (\ref{uPDF2.1g})
as well as the general formula~(\ref{eq-KMR-int}) with CT'14 set
results in a good (within the estimated uncertainties) description of $b\bar b$-dijet cross sections
reported by the ATLAS Collaboration, as it is demonstrated in Fig.~\ref{fig5}.
Such events provide us with additional observables which are a useful complementary ones.
%when studying the heavy flavor production.
In particular, good agreement of our predictions and measured
$\Delta \phi$ distributions is remarkable,
where $\Delta \phi$ is the difference between the azimuthal angles of the produced $b$-jets.
The latter is known to be strongly sensitive to the
high-order pQCD corrections encoded in the
$\mathbf{k}_T^2$ shape of the TMD parton densities (see\cite{bb-kt-our1, bb-kt-our2}).
In fact, $\Delta \phi$ distribution
degenerates into $\delta$-function at $\phi = 0$ at the LO pQCD
and continuous spectra can only be obtained by including higher-order corrections,
which are automatically taken into account in our calculations in the form of TMD parton densities in a proton.

Another interesting observable in $b\bar b$-dijet production
is the angular variable $\chi$. % = \exp |y_1 - y_2|$.
This variable is constructed in a way
that the $2 \to 2$ cross section of elastic scattering of two point-like
massless particles is approximately constant as a function of
$\chi \sim (1 + \cos \theta^*)/(1 - \cos\theta^*)$, where $\theta^*$
is the scattering angle in the center-of-mass frame.
It was argued\cite{bjet-ATLAS-7} that this observable
is relatively insensitive to the uncertainties
related with the PDFs behaviour at small $x$.
One can see that the $\chi$ distribution is well reproduced by our
theoretical calculations based on both KMR/WMR definitions.
The distribution flattens for large invariant mass $M$.

Finally, we conclude that the newly derived analytical expressions for the TMD
parton densities in a proton, valid in the whole $x$ range,
agree well with
%does not contradict
available LHC data on heavy flavor
production in $pp$ collisions. The integral definition of the KMR/WMR
approach (\ref{eq1-Sg_kt}) --- (\ref{uPDF2.1g}) provides a better description of the data.
The corresponding TMD gluon distribution in a proton, labelled as KL'2025, is already available for
public usage and implemented into the \textsc{tmdlib} package\cite{TMDLib2} and
Monte-Carlo event generator \textsc{pegasus}\cite{PEGASUS}.
It supersedes our previous KLSZ'2020 set.

\section{Conclusion} \indent

We presented the analytical expressions for the Transverse Momentum Dependent parton densities in a proton.
These expressions are derived at leading order of the QCD running coupling
and valid at both small and large $x$.
The calculations are performed
using the Kimber-Martin-Ryskin/Watt-Martin-Ryskin approach
with different treatment of kinematical constraint, which reflects the angular and
strong ordering conditions for parton emissions
at the last step of the QCD evolution.
Both the differential and integral definitions of the KMR/WMR scenario
are considered.
As an input, the analytical solution of DGLAP equations
for conventional parton distributions in a proton is applied,
where the valence and non-singlet quark parts obey the Gross-Llewellyn-Smith and
Gottfried sum rules and momentum conservation for the singlet quark and gluon densities is taken into account.

We demonstrated that the derived expressions
%does not contradict
agree well with
the LHC data on inclusive heavy flavor production
in $pp$ collisions at high energies.
It was confirmed that the integral definition
of the KMR/WMR approach provides better description of the data
and more preferable in the phenomenological applications.
The corresponding TMD gluon distribution in a proton, labelled as KL'2025, is already available for
public usage and implemented into the \textsc{tmdlib} package and
Monte-Carlo event generator \textsc{pegasus}. It supersedes our previous KLSZ'2020 set.

The main advantage of developed approach is related with quite
compact analytical formulas for the TMD parton densities in a proton, which could be
easily applied in further phenomenological studies.
Of course, the presented LO analysis can be considered as one of the first steps in
our investigations in this direction. So, we are going to improve the accuracy to the NLO level in
fortcoming studies. Also, we plan to extend the consideration to nuclear PDFs and
nuclear TMDs (following \cite{nTMDs-KMR-our,EMC-lowx-our}) by using the results for the whole $x$ range presented here).
It is important for future studies of
proton-nucleus and nucleus-nucleus interactions within the TMD-based approaches.

\section*{Acknowledgements} \indent

We thank S.P.~Baranov, M.A.~Malyshev and H.~Jung for their
interest, very important comments and remarks.
%A.V.L. and A.V.K. would like to thank School of Physics and
%Astronomy, Sun Yat-sen University (Zhuhai, China) for extremely warm hospitality.
Our study was carried out at the expence of the Russian Science Foundation
grant 25-2200066.

\bibliography{KMR}

\end{document}